\documentclass[11pt,preprint]{aastex}

\begin{document}

\title{Spatial Growth of the Current-Driven Instability in Relativistic Jets}

\author{
Yosuke Mizuno\altaffilmark{1},  Philip E. Hardee\altaffilmark{2}, and Ken-Ichi Nishikawa\altaffilmark{3} }

\altaffiltext{1}{Institute of Astronomy, National Tsing-Hua University, Hsinchu 30013, Taiwan, R.O.C.; mizuno@phys.nthu.edu.tw}
\altaffiltext{2}{Department of Physics and Astronomy, The University of Alabama, Tuscaloosa, AL
35487, USA}
\altaffiltext{3}{Department of Physics, University of Alabama in Huntsville, Huntsville, AL 35899, USA}

\shorttitle{3DRMHD Simulations of CDI in Relativistic Jets}
\shortauthors{Mizuno et al.}

\begin{abstract}
We have investigated the influence of velocity shear and a radial density profile on the spatial development of the current driven kink instability along helically magnetized relativistic jets via three-dimensional relativistic magnetohydrodynamic simulations. In this study, we use a non-periodic computational box, the jet flow is initially established across the computational grid, and a precessional perturbation at the inlet triggers growth of the kink instability.  If the velocity shear radius is located inside the characteristic radius of the helical magnetic field, a static non-propagating current driven kink is excited as the perturbation propagates down the jet. Temporal growth disrupts the initial flow across the computational grid not too far from the inlet. On the other hand, if the velocity shear radius is outside the characteristic radius of the helical magnetic field, the kink is advected with the flow and grows spatially down the jet. In this case flow is maintained to much larger distances from the inlet. The effect of different radial density profiles is more subtle. When the density increases with radius, the kink appears to saturate by the end of the simulation without apparent disruption of the helical twist. This behavior suggests that relativistic jets consisting of a tenuous spine surrounded by a denser medium with a velocity shear radius outside the radius of maximum toroidal magnetic field have a relatively stable configuration.
\end{abstract}

\section{Introduction}

Relativistic jets are ubiquitous features of many accreting black hole and neutron star systems. They are associated with X-ray binaries (XRBs) and Gamma-Ray Bursts (GRBs) containing neutron stars or stellar-mass black holes (BHs) of mass $M \sim 1.4 - 20~M_{\odot}$, and active galactic nuclei (AGNs) with super massive BHs of mass $M \sim 10^6 - 10^{10}~M_{\odot}$.

It is commonly believed that jets are powered and collimated by magnetohydrodynamic (MHD) processes (Lovelace 1976; Blandford 1976, 2000; Blandford \& Znajek 1977). General relativistic magnetohydrodynamic (GRMHD) simulations of jet formation (e.g., De Villiers et al. 2003, 2005; Hawley \& Krolik 2006, McKinney \& Gammie 2004; McKinney 2006; McKinney \& Blandford 2009; Beckwith et al. 2008; Hardee et al. 2007; Komissarov \& Barkov 2009; Penna et al. 2010) show development of turbulence via magneto-rotational instability (MRI) (Balbus \& Hawley 1998) and angular momentum transfer in the accretion disk, leading to diffusion of matter and magnetic field inwards, and generation of unsteady outflows. In general, GRMHD simulations with spinning black holes indicate jet production consisting of a Poynting dominated (in the sense that the energy is transferred predominantly by the electromagnetic field) high Lorentz factor spine, and a matter dominated, mildly relativistic sheath with $v < c$ possibly embedded in a sub-relativistic speed, $v \ll c$, disk/coronal wind.  Note however, the that the Lorentz factor can achieve a maximum in a tenuous boundary layer between a Poynting-flux spine and a dense lower speed kinetically dominated sheath via rarefaction acceleration (e.g, Aloy \& Rezzolla 2006; Mizuno et al. 2008; Zenitani et al. 2010a; Komissarov et al. 2010). 

It is thought that relativistic jets on the largest scales are kinetically dominated, i.e., most of the electromagnetic energy is converted to kinetic energy. The means by which magnetic energy is converted to kinetic energy has not been identified, but it is generally thought to involve gradual acceleration by magnetic forces (Beskin \& Nokhrina 2006; Komissarov et al 2007, 2009, 2010; Tchekhovskoy et al. 2009, 2010; Lyubarsky 2009, 2010a, 2011; Granot et al. 2011; Lyutikov 2011; Granot 2012a,b) and/or non-ideal MHD effects, specifically magnetic reconnection (Spruit et al. 2001; Drenkhan 2002, Drenkhahn \& Spruit 2002;  Lyubarsky 2010b; McKinney \& Uzdensky 2012). The large scale magnetic field may dissipate if the regular magnetic structure is destroyed as a result of a global MHD instability, the kink instability being the most plausible candidate (Lyubarskii 1992, 1999; Eichler 1993; Spruit et al 1997; Begelman 1998; Giannios \& Spruit 2006).

The most promising MHD models for the acceleration and collimation of jets involve the presence of a magnetic field with foot points anchored to a rotating object (an accretion disk or a spinning neutron star or black hole). The dominance of the toroidal component ($B_\phi$) over the poloidal component ($B_p$) is a natural consequence in these models. It is well known that current carrying plasma columns containing strong toroidal magnetic fields are unstable to non-axisymmetric perturbations (Bateman 1978). Among these current-driven ({\bf CD}) instabilities, the kink mode is the most violent. The kink mode leads to helical displacement of the plasma-column and may disrupt the system. Thus, non-linear development of the kink mode could trigger violent magnetic dissipation in relativistic outflows. McKinney \& Blandford (2009) pointed out that  when the Kruskal-Shafranov (KS) instability criterion, $|B_{p}/B_{\phi}| < z / 2 \pi R$, for cylindrical force-free equilibria is applied to relativistic jets, the jets become unstable beyond the Alfv\'en surface located at $z \simeq 10 R_g$ where $R_g$ is the gravitational radius, and where the jet radius $R_j \lesssim z$ and  $|B_\phi| \gtrsim |B_p|$. However, jet rotation and velocity shear significantly affect the instability criterion (Istomin \& Pariev 1994, 1996;  Lyubarskii 1999; Tomimatsu et al. 2001; Narayan et al. 2009) and need to be studied in this context.  

On the smallest scales, the rapid observed variability of X-ray/TeV gamma-ray flares with timescales from several minutes to a few hours in blazars pose severe constraints on the particle acceleration timescale and the emission region size and also provide indirect evidence for instabilities. Here particle acceleration provided by current driven instability and magnetic reconnection conversion of Poynting flux to kinetic flux (Sikora et al. 2005) to generate flares would have to come from small, a few Schwarzschild radii in size, fast moving ``needles" or ``jet" within a slower jet medium, the so-called ``needles or jet-in-a-jet" scenarios (Levinson 2007; Begelman et al. 2008; Ghisellini \& Tavecchio 2008; Giannios et al. 2009). 

On somewhat larger scales, high resolution VLBI observations show that many AGN jets display helical structures on sub parsec to kiloparsec scales (e.g., G\'{o}mez et al. 2001; Lobanov \& Zensus 2001) that provide indirect evidence for instability. Helical distortion may be caused by precession of the jet and/or by MHD instabilities (current-driven or Kelvin-Helmholtz velocity shear driven). Non-relativistic and relativistic simulations of magnetized jet formation and/or propagation have showed helical structures attributed to CD kink instability (e.g., Lery et al. 2000; Ouyed et al. 2003; Nakamura \& Meier 2004; Nakamura et al. 2007; Moll et al. 2008; Moll 2009; McKinney \& Blandford 2009; Carey \& Sovinec 2009; Mignone et al. 2010; Porth 2013; Guan et al. 2013). In the absence of CD kink instability and resistive relaxation, helical structures may be attributed to the Kelvin-Helmholtz ({\bf KH}) helical instability driven by velocity shear at the boundary between the jet and the surrounding medium (e.g., Hardee 2004, 2007; Mizuno et al. 2007) or triggered by precession of the jet ejection axis (Begelman et al. 1980). It is still not clear whether CD, velocity shear driven, or jet precession is responsible for the observed structures, or whether these different processes are responsible for the observed twisted structures at different spatial scales.

In previous work, we investigated the temporal development of the CD kink instability using a periodic computational box. In Mizuno et al. (2009), we studied the instability of helically magnetized static plasma columns (or more generally rigidly moving flows considered in the proper reference frame). Simulation results showed that the initial configuration was strongly distorted but not disrupted by the CD kink instability. The growth rate depended on the radial profile of the magnetic pitch and the density. The plasma column was more unstable when the helicity of the magnetic field lines increased with radius (magnetic pitch parameter, $P \equiv RB_z/B_{\phi}$, decreased with radius $R$) and when the density decreased with radius. In Mizuno et al. (2011), we investigated the influence of velocity shear on the CD kink instability and found that the kink propagates along the jet with speed and flow structure dependent on the radius of the velocity shear (hereafter called the velocity shear radius) relative to the characteristic radius of the helically twisted force-free magnetic field. In Mizuno et al. (2012), we studied the influence of jet rotation and differential motion on the CD kink instability, and in accordance with linear stability theory, we found that development of the instability depends on the lateral distribution of the poloidal magnetic field. When the profile of the poloidal field is shallow, the instability develops slowly and eventually saturates. When the profile of the poloidal field was steep and the magnetic pitch in the inner portion of the jet was sufficiently small, i.e., the magnetic field was sufficiently tightly twisted, multiple growing helical wavelengths disrupted the initial configuration.  Recently O'Neill et al. (2012) have also investigated the CD kink instability in a local, co-moving frame using the 3D RMHD module within the Athena code (Beckwith \& Stone 2011). They also studied rotating jets with a purely toroidal magnetic field, the hoop stress being balanced by the gas pressure and centrifugal force, and found that the instability brings the system into a turbulent state. All of these results have been obtained using a periodic computational box that allows study of the temporal growth of a few axial wavelengths that fit within the small simulation domain.  The time has come to begin systematic more physically realistic simulations that study the spatial development of CD instability.

In order to investigate the spatial development of the CD kink instability a non-periodic computational box must be used. At this time, only a few 3D simulations of relativistic, magnetized jets have been performed that study the spatial properties of the CD kink instability, and the results are still controversial. McKinney \& Blandford (2009) simulated the generation and propagation of a relativistic highly magnetized jet and found no significant development of non-axisymmetric instabilities, even though the simulation satisfied the KS criterion for instability. On the contrary, Mignone et al. (2010) found strong jet distortion in their jet simulation. The discrepancy may be attributed to the difference in the setup. Mignone et al. (2010) assumed that the magnetic field was purely toroidal, whereas McKinney \& Blandford (2009) simulated jet launching from a rotating magnetized configuration and their jet contained both toroidal and poloidal magnetic field components. Recently, Porth (2013) performed a simulation of relativistic jet formation from magnetospheres in Keplerian rotation. The simulation showed development of the kink mode but the kink saturated before notable dissipation or disruption occurred. 

In this paper, we begin a systematic study of the spatial development of the CD kink instability in a relativistic jet via 3D RMHD simulations using a non-periodic computational box. In order to isolate the role of different factors and therefore to gain physical insight we do not attempt to properly account for poorly known relativistic jet poloidal and radial profiles. Here we concentrate on the role of the velocity shear radius relative to the radius at which the toroidal component of the magnetic field is maximum and the radial density profile. We do not investigate the effect of a radial velocity or magnetic pitch profile in this study. We know that the temporal development of the CD kink instability is dependent on differential motion, the magnetic pitch profile, and the density profile (Appl et al. 2000; Mizuno et al. 2009, 2011, 2012).  Our previous temporal growth simulations (Mizuno et al. 2009) considered declining and constant density profiles with constant density profiles being slightly more stable. In order to compare spatial results to previous temporal results and to more closely simulate a high speed tenuous jet surrounded by a denser sheath or wind, we consider declining, constant and increasing density profiles. Since temporal stability results (Mizuno et al. 2009, 2011, 2012) indicate more substantial instability dependence on the magnetic pitch and velocity profiles that are unknown for observed relativistic jets, a choice of constant magnetic pitch and sharp velocity shear is reasonable for this first study. 

We describe the numerical method and setup used for our simulations in Section 2, present our results in Section 3, and in Section 4 compare our results to instability expectations and conclude.

\section{Numerical Method and Setup}

We perform RMHD simulations of the CD kink instability using the 3D GRMHD code ``RAISHIN"  (Mizuno et al. 2006, 2011). The code setup for spatial development of the CD kink instability is similar to the setup for temporal development in a periodic computational box outlined in Mizuno et al. (2011).  Here we use a non-periodic simulation box that is longer in the jet propagation direction. The computational domain is $6L \times 6L \times 20L$  in a Cartesian ($x, y, z$) coordinate system with grid resolution of $\Delta L = L/40$ for the transverse $x$ and $y$-directions and $\Delta L = L/20$ for the axial $z$-direction. We impose outflow boundary conditions on all surfaces except the inflow plane at $z = 0$.  In the simulations, a preexisting jet flow is established across the computational domain. This setup represents the case in which the jet is in equilibrium with an external medium far behind the leading-edge. 

As in Mizuno et al. (2011), we choose a force-free helical magnetic field for the initial configuration (Mizuno et al. 2009). A force-free configuration is a reasonable choice for a magnetically dominated jet. Following previous work, we choose poloidal and toroidal magnetic field components written in the following form
\begin{equation}
B_{z}= {B_{0} \over [1+ (R/a)^{2}]^{\alpha}}~,
\end{equation}
\begin{equation}
B_{\phi}= {B_{0} \over (R/a)[1+ (R/a)^{2}]^{\alpha}} \sqrt{ { [1 +
(R/a)^{2}]^{2 \alpha} -1 - 2 \alpha (R/a)^2 \over 2 \alpha -1}}~,
\end{equation}
where $R$ is the radial position in cylindrical coordinates, $B_{0}$ parameterizes the magnetic field, $a$ is the characteristic radius of the magnetic field (the toroidal field component is a maximum at $a$ for constant magnetic pitch), and $\alpha$ is a pitch profile parameter. The radial profile of the magnetic pitch, $P \equiv R B_{z} / B_{\phi}$, is determined  by the pitch parameter $\alpha$, where smaller $P$ indicates increased helicity of the magnetic field lines. In the present simulations, we choose a characteristic magnetic radius of $a=L/4$ (the computational domain is $24a \times 24a \times 80a$), and we choose $\alpha=1$ which gives constant magnetic pitch and magnetic helicity.
 
We consider a low gas pressure medium with pressure decreasing radially proportional to $a/R$ when $R \ge a$, where $p = p_{0} =0.01$ in units of $\rho_{0}c^{2}$ is constant for $R < a$ (Mizuno et al. 2012).  We choose this radial gas pressure decline in order to keep the sound speed from exceeding the jet speed. This choice for the gas pressure means that we are not strictly in a static steady state equilibrium initial configuration.  We choose three different radial density profiles: (A) decreasing, (B) constant and (C) increasing with radius that allow us to study the impact of different radially decreasing Alfv\'en speeds on spatial growth. The decreasing density profile is given by $\rho = \rho_1 \sqrt{B^2/B_0^2}$ and the increasing density profile is given by $\rho=\rho_1 \sqrt{B_0^2/B^2}$ where $\rho_1=0.8 \rho_0$. The magnetic field amplitude is $B_0 = 0.4$ in units of  $\sqrt{4\pi\rho_{0}c^{2}}$ leading to a low plasma $\beta \equiv P_{gas}/P_{mag}$ near the axis. The equation of state is that of an ideal gas with $p=(\Gamma -1) \rho e$, where $e$ is the specific internal energy density and the adiabatic index $\Gamma=5/3$. The specific enthalpy is $h \equiv 1+e/c^{2} +p/\rho c^{2}$.  The sound speed is $c_{s}/c \equiv (\Gamma p/\rho h)^{1/2}$ and the Alfv\'{e}n speed is given by $v_{A}/c  \equiv [b^{2}/(\rho h +b^{2})]^{1/2}$, where $b$ is the magnetic field measured in the comoving frame, $b^{2}=\mathbf{B}^{2}/\gamma^{2}+(\mathbf{v} \cdot \mathbf{B})^{2}$ (Komissarov 1997; Del Zanna et al. 2007). The magnetosonic speed (e.g., Vlahakis \& K\"onigl 2003; Hardee 2007) is given by $v_{ms} \equiv (c_s^2 + v_A^2 - c_s^2v_A^2/c^2)^{1/2} = (c_s^2/\gamma_A^2 + v_A^2)^{1/2}$ where $\gamma_A \equiv (1 - v_A^2/c^2)^{-1/2}$. The sound, Alfv\'en and magnetosonic speeds on the axis are $c_{s0} = 0.14c$, $v_{A0} = 0.40c$, and $v_{ms0} = 0.42c$. 

We choose the jet to be mildly-relativistic, $v_{j}=0.2c$. This choice allows us to compare our spatial results to our previous temporal results (Mizuno et al.\ 2011) and other previous Newtonian simulations (Nakamura \& Meier 2004; Moll et al. 2008; Moll 2009).  We perform simulations with velocity shear radii of $R_{j}= L/8$ ($a/2$) and $L$ ($4 a$) at which a sharp transition from jet  with speed $v=v_j$ to stationary ambient medium with $v = 0$ occurs (see Fig.\ 1d). With these two choices the velocity shear is either well inside or well outside the characteristic radius of the magnetic field.
These two choices will allow us to investigate the dependence of kink propagation on location of the velocity shear radius relative to the characteristic radius of the magnetic field, and also to compare our spatial results  to previous temporal results (Mizuno et al. 2012). To break the equilibrium state, a precessional perturbation is applied at the inflow using a transverse velocity component with magnitude $v_\perp = 0.01 v_j = 0.002c$, and an angular frequency $\omega_j  = 0.42$~c/L. We define a characteristic wavelength $\lambda_j \equiv 2 \pi v_j / \omega_j \simeq 3L =12a$. 

The initial radial profiles of the magnetic field, density, sound speed and Alfv\'{e}n speed are shown in Figure 1. We find that these initial radial profiles are only minimally modified by the radially decreasing gas pressure and the minimal modification to the initial radial profiles observed at very early times in the simulations is overwhelmed at early times in the simulations by changes resulting from the precessional perturbation.  
\begin{figure}[h!]
\epsscale{0.8}
\plotone{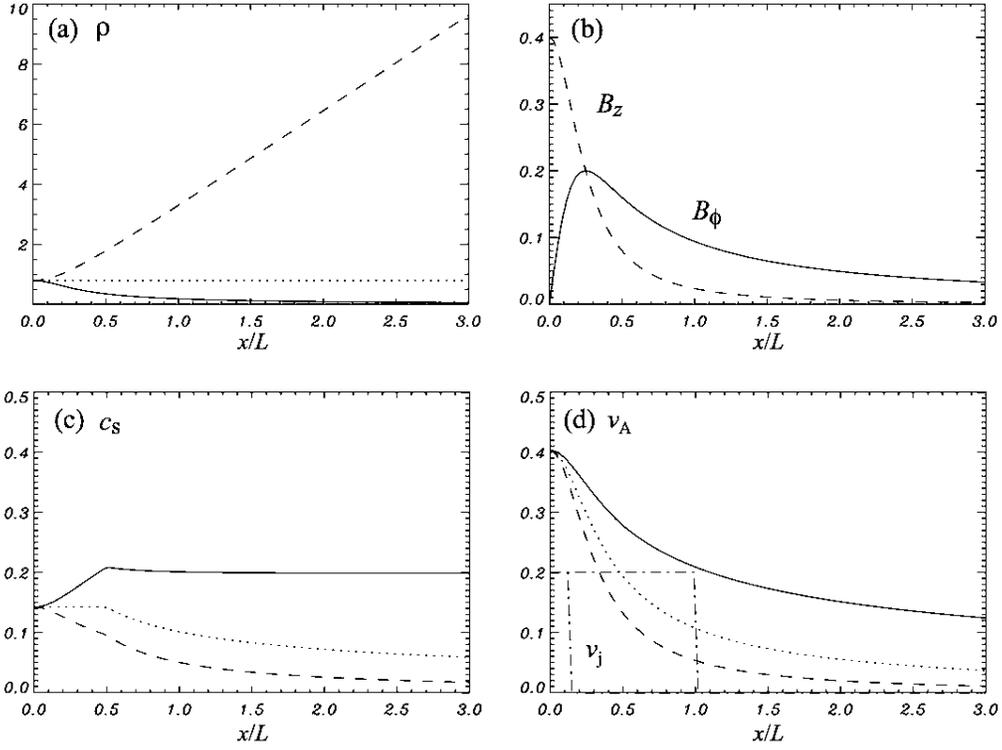}
\caption{Radial profiles of (a) the density, (b) the toroidal ($B_{\phi}$, solid) and axial ($B_{z}$, dashed) components of the magnetic field, (c) the sound speed $c_s/c$, and (d) the Alfv\'{e}n speed $v_A/c$. In panels a, c \& d, the (A) decreasing, (B) constant and (C) increasing radial density profile cases are indicated by the solid, dotted, and dashed lines, respectively. The dash-dotted lines in panel d indicate the radial location of the velocity shear radii $R_j=a/2$ and $4a$ and the jet speed $v_j/c=0.2$.   
\label{f1}}
\end{figure}
Our choice for the jet velocity and radial density profiles keeps the jet flow sub-Alfv\'{e}nic for decreasing density cases (A1 \& A2) with $R_{j}= a/2$ and $4 a$, and for the constant density case (B1) and increasing density case (C1) with $R_{j}= a/2$.  For the constant density case (B2) and increasing density case (C2) with $R_{j}= 4 a$ the jet is sub-Alfv\'enic in the innermost regions and becomes super-Alfv\'enic inside the velocity shear radius. In general, the jet speed is mildly supersonic in the innermost regions for all cases, more supersonic in the outer regions for constant (B) and increasing (C) density cases, and transonic in the outer regions for the decreasing (A) density cases. Here we focus primarily on the development of the CD kink instability, although note that the constant (B2) and increasing (C2) density cases  with $R_{j}= 4 a$ are weakly super-magnetosonic inside and outside the velocity shear radius and thus potentially velocity shear driven Kelvin-Helmholtz (KH) unstable. The various different cases and model parameters at selected radial locations are summarized in Table 1.
\begin{deluxetable}{cccccccccccc}
\tablecolumns{12}
\tablewidth{0pc}
\tablecaption{Models and Parameters}
\label{table1}
\tablehead{
\colhead{Case} & \colhead{$R_{j}/a$} & \colhead{$\rho$} & \colhead{$c_{sa/2}$\tablenotemark{\,a}} & \colhead{$v_{Aa/2}$} & 
\colhead{$v_{msa/2}$}  & \colhead{$c_{sa}$\tablenotemark{\,b}} & \colhead{$v_{Aa}$} & 
\colhead{$v_{msa}$} & \colhead{$c_{s4a}$\tablenotemark{\,c}} & \colhead{$v_{A4a}$} & 
\colhead{$v_{ms4a}$}} 
\startdata
A1 & 0.5 & $\propto \rho_1B$ & 0.15 & 0.38 & 0.41 &  0.17 & 0.34 & 0.38 & 0.20 & 0.21 & 0.29 \\
A2 & 4.0  &  '' & '' & '' & '' & '' &   '' & '' & '' & '' & '' \\   
B1 & 0.5   & $= \rho_1$ & 0.14 & 0.37 & 0.39 &  0.14 & 0.29 & 0.32 & 0.10 & 0.11 & 0.15\\
B2 & 4.0   & '' & '' & '' & '' & '' &  '' & '' & '' & '' & ''  \\
C1 & 0.5 & $\propto \rho_1/B$ & 0.12 & 0.25 & 0.27 &  0.13 & 0.35 & 0.37 & 0.050 & 0.053 & 0.073 \\
C2 & 4.0  & '' & '' & '' & '' & '' &  '' & '' & '' & '' & ''  \\   
\enddata
\tablenotetext{\,a}{~Values calculated at $R = a/2$}
\tablenotetext{\,b}{~Values calculated at $R = a$}
\tablenotetext{\,c}{~Values calculated at $R = 4a$}
\end{deluxetable}
\clearpage
 
\section{Simulation Results}

\subsection{Global Kink and Flow Structure}

The temporal and spatial development of the CD kink instability is shown for cases (A) and (C) in Figures 2 and 3, respectively.  Here we show only the decreasing (A) and increasing (C) density cases as only these cases are needed to illustrate the density dependence. The figures show  3D density isosurfaces at two different simulation times where the simulation time $t$ is in units of $t_c \equiv L/c = 4a/c$.
For both velocity shear radii  the precessional perturbation leads to growth of the CD kink instability and the magnetic field lines wind around the developing helical density structure. While not easily seen in the density isosurfaces, at longer times shorter and longer CD kink wavelengths develop at smaller and larger radii, respectively.
\begin{figure}[h!]
\epsscale{0.75}
\plotone{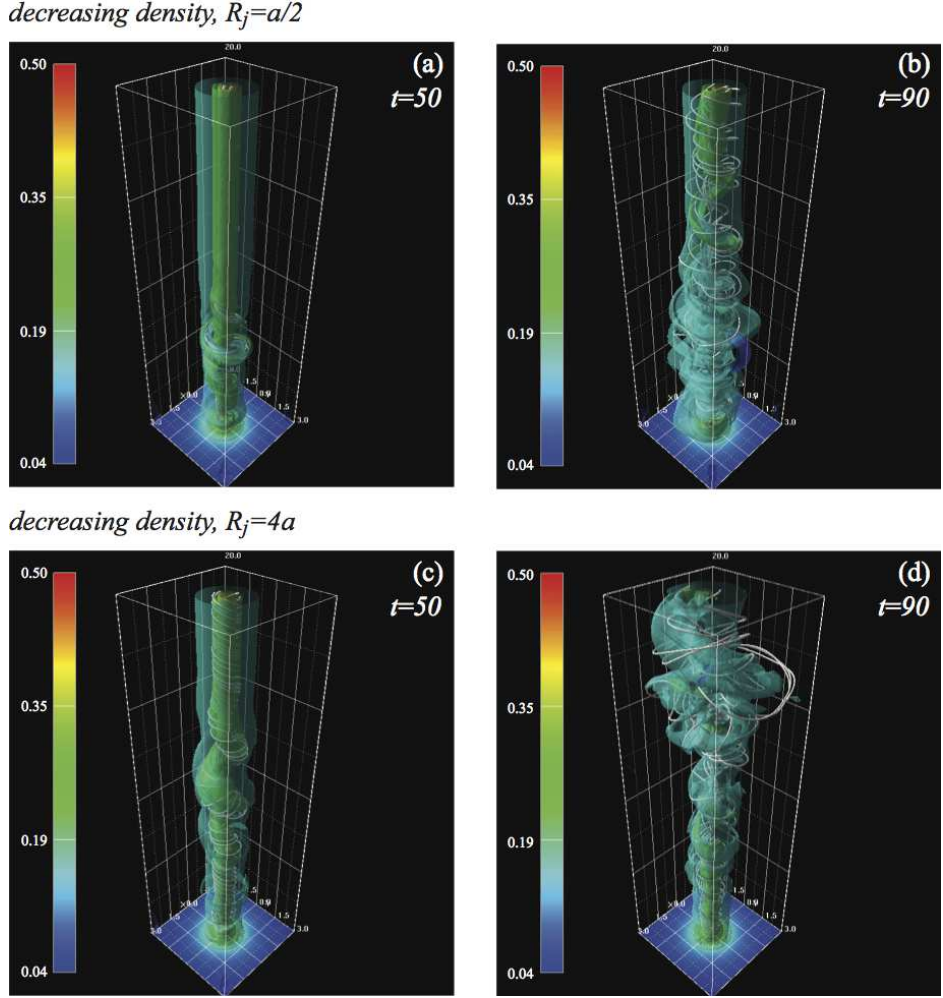}
\caption{Three-dimensional density isosurfaces with a transverse slice at $z=0$ for the decreasing density cases B1 ($R_j=a/2$) and B2 ($R_j=4a$) at $t/t_c =50$ and $90$. Representative solid magnetic field lines are shown and color scales with the logarithm of the density. \label{decd_3D}}
\end{figure}

For the declining density cases, Figure 2 at simulation time $t = 50$ shows that for velocity shear radius $R_j=a/2$ (case A1), the helical kink develops rapidly near the jet inlet and at this time the largest amplitudes are at $z \lesssim 5L$. At the later simulation time, $t = 90$, the kink amplitude continues to grow near the jet inlet and also develops farther down the jet. The largest amplitudes are found at $z \sim 5-7 L = 20 - 28a$. Note that jet plasma flowing at $v_j = 0.2c$ travels across the computational grid in time $t = 100 t_c$, i.e., $20L = 80a = 100 v_j t_c$. This result indicates that the perturbation propagates down the jet to excite the CD kink instability which grows temporally in place to achieve the largest amplitudes near to the inlet.  At larger distances from the inlet amplitudes decline as there has been less time for the kink to grow after the perturbation induced at the inlet passes. The amount of time after the perturbation has passed a given location depends on the propagation speed of the perturbation. We find that the perturbation propagation speed is greater than the flow speed (see Section 3.2) and the precessional perturbation crosses the grid before $t=90$.

Figure 2 shows that for velocity shear radius $R_j=4a$ (case A2), the helical kink again first develops near to the inlet and then develops farther down the jet as the simulation time increases.  Even at simulation time $t=50$ results suggest spatial growth with the largest amplitudes at $z \sim 7-9 L = 28 - 36a$, considerably beyond where the largest amplitudes where found for case A1 even at the later simulation time.  For case A2 at the later simulation time, $t =90$, it is clear that the kink amplitude increases spatially down the jet with the largest amplitudes located far from the inlet at $z \sim 15L = 60a$.  The very different result from the $R_j=a/2$ case indicates a propagating spatially growing kink as opposed to a static temporally growing kink.  We find that the perturbation propagation speed is greater than the flow speed (see Section 3.2) and the precessional perturbation crosses the grid before $t=90$.  The difference in behavior between case A1 and A2 was suggested by previous periodic box simulations of temporal kink growth  which showed a temporally growing static kink when $R_j=a/2$ and a temporally growing moving kink when $R_j=4a$ (Mizuno et al. 2011). In both decreasing density cases, organized helical density  and magnetic structure appear disrupted at the terminal simulation time, albeit at very different distances from the inlet.

For the increasing density cases, Figure 3 reveals some differences relative to the comparable declining density cases.  
\begin{figure}[h!]
\epsscale{0.75}
\plotone{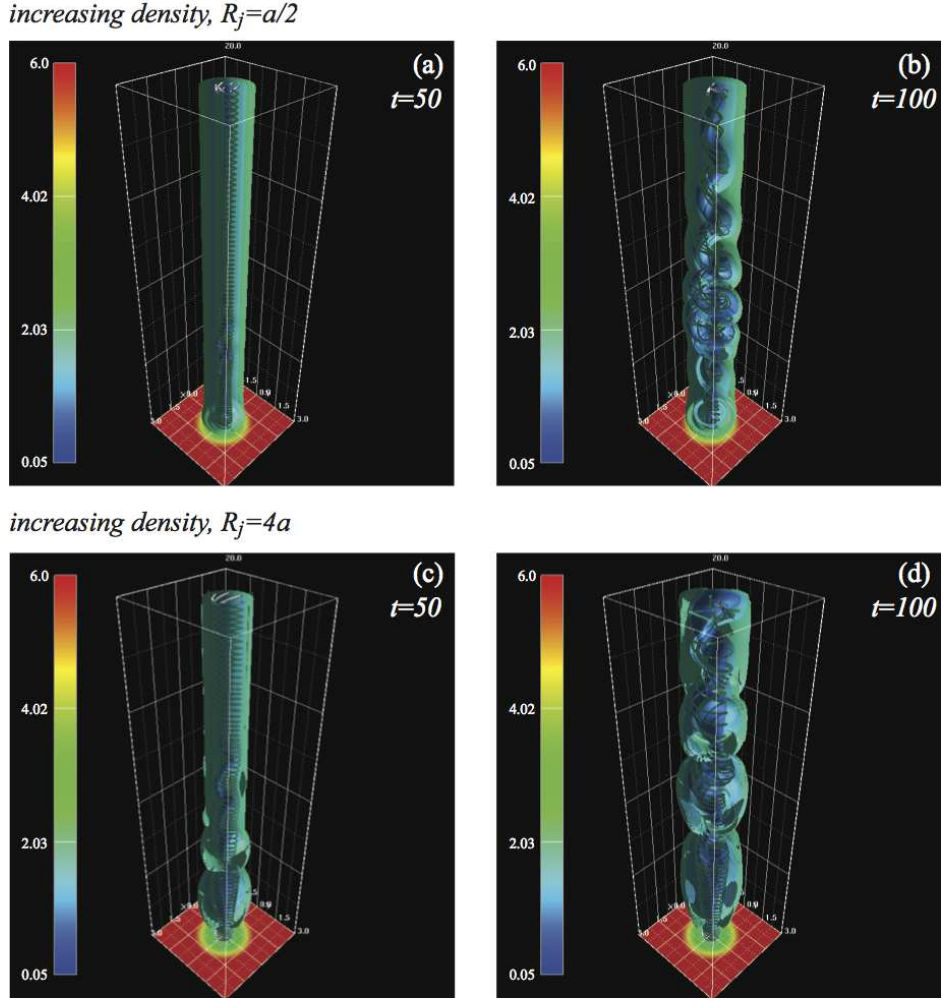}
\caption{Three-dimensional density isosurfaces with a transverse slice at $z=0$ for the increasing density cases C1 ($R_j=1/2a$) and C2 ($R_j=4a$) at $t/t_c =50$ and $100$. Representative solid magnetic field lines are shown and the color scales with the logarithm of the density.
 \label{incd_3D}}
\end{figure}
Like the declining density cases A1 and A2 at $t = 50$, in cases C1 and C2 the helical kink develops first near the jet inlet. At the later simulation time, now $t = 100$, in case C1 with $R_j = a/2$ the kink amplitude continues to grow near the jet inlet and also develops farther down the jet like the decreasing density case A1. Now the largest amplitudes are found somewhat farther from the inlet at $z \sim 7.5L = 30a$ and appear less than for the comparable decreasing density case A1.  For $R_j=4a$, at the later simulation time the kink amplitude increases spatially down the jet like the decreasing density case A2. The amplitude appears to saturate beyond about $z \sim 10L = 40a$ without obvious disruption of the helical twist. However, beyond $z \sim 15L = 60a$ the magnetic structure becomes more distorted and less regular. We note that the perturbation propagation speed is greater than the flow speed (see Section 3.2) and the precessional perturbation crosses the grid before $t=100$.
Comparison between the two extreme density profile cases shows faster temporal and/or spatial growth for the decreasing density cases. This result is expected as temporal growth was more rapid on the periodic grid for decreasing density cases relative to constant density cases and this would convert to more rapid spatial growth if the kink is propagating.  The reason for more rapid growth in the decreasing density cases results from the higher Alfv\'{e}n speed far from jet axis which results in a higher growth rate (Appl et al. 2000; Mizuno et al. 2008). 

In Figure 4  we show 3D total velocity isosurfaces near the end of the decreasing density and increasing density simulations in order to better understand the effect of the temporally and/or spatially growing kink on jet flow and magnetic field structure.
Here the distance to which the jet flow remains collimated and helical magnetic structure remains organized is better revealed.  
\begin{figure}[h!]
\epsscale{0.75}
\plotone{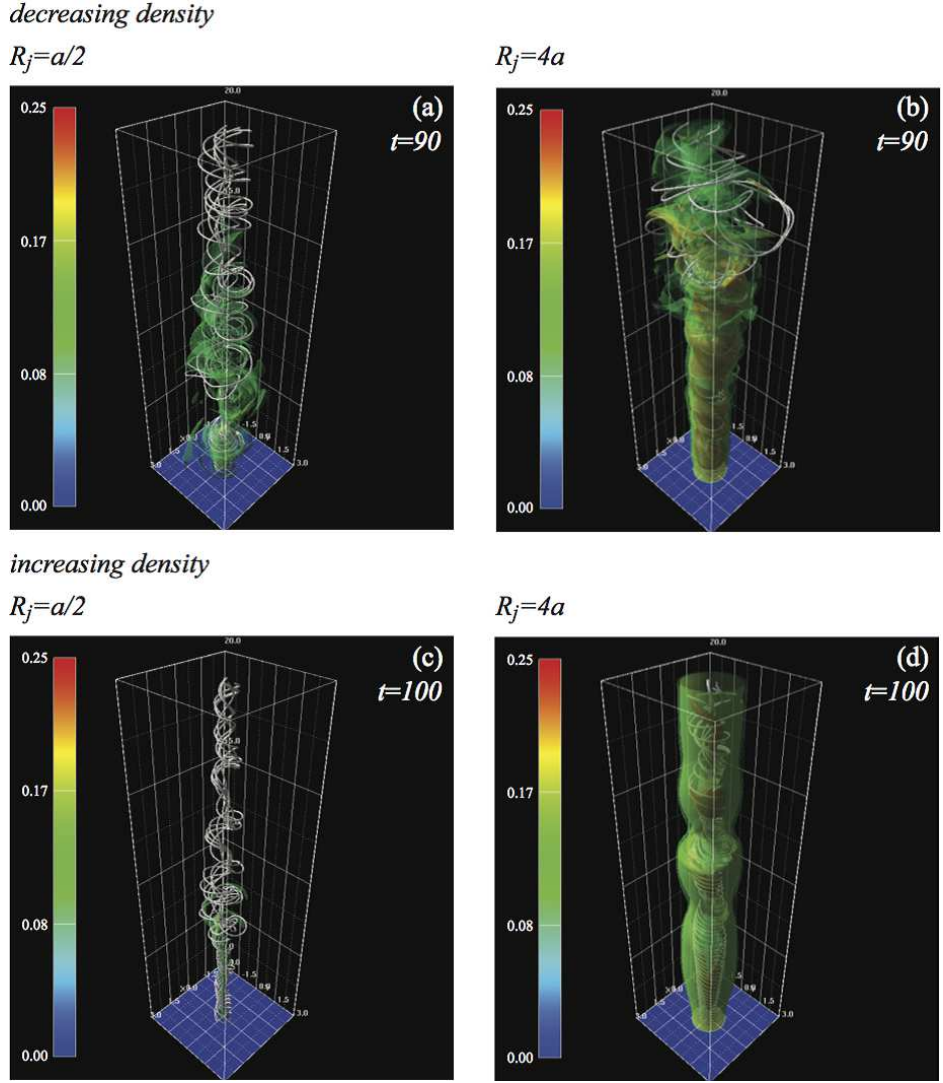}
\caption{Three-dimensional total velocity isosurfaces for $v_t \ge 0.1 c$ where $v_t = \sqrt{(v_x^2 + v_y^2 + v_z^2}$ is indicated by the color scale with a transverse slice at $z=0$. The decreasing density cases (A1 \& A2) are shown in the upper panels (a) $R_j = a/2$ and (b) $R_j = 4a$, and the increasing density cases (C1 \& C2) are shown in the lower panels (c) $R_j = a/2$ and (d) $R_j = 4a$ with representative solid magnetic field lines. \label{vtot_3D}}
\end{figure}

For the decreasing density case A1 with $R_j = a/2$ both flow and magnetic field structure are highly distorted beyond $z \sim 3L$. Isosurface flow speeds greater than 0.1c are seen out to $z \sim 15L$ and indicate helical twisting of the flow.  For the decreasing density case A2 with $R_j = 4a$ both the flow and magnetic field structure are highly distorted beyond $z \sim 10L$ although isosurface flow speeds greater than 0.1c are seen all the way to $z = 20L$.  We conclude that both the regular helical magnetic field structure and highly collimated flow structure are disrupted on the computational grid for these two cases although case A2 where the CD kink is advected with the flow maintains a collimated flow and organized helical magnetic field to considerably larger distance.  
For the increasing density case C1 with $R_j = a/2$ both flow and magnetic field structure are highly distorted beyond about $z \sim 5L$ and in this case isosurface flow speeds greater than 0.1c are seen only to $z \sim 7.5L$. As in the comparable decreasing density case A1 the velocity isosurface indicates helical twisting of the flow inside $z \sim 7.5L$. For the increasing density case C2  with $R_j = 4a$ flow collimation and helical magnetic structure appear maintained across the computational grid albeit with significant twisting beyond about $z \sim 10L$. Note the helical twist in the inner high speed flow field (red velocity isosurface) markedly apparent beyond about $z \sim 15L$.

Comparison between the decreasing and increasing density profile cases, indicates that the helical magnetic field maintains a more regular structure to larger distances for increasing density profile cases C1 and C2 when compared to the comparable decreasing density profile cases A1 and A2. The initial flow maintains a more regular structure to larger distances for the increasing density profile case C2 when compared to the comparable density profile case A2.
In the 3D velocity isosurface images higher total velocities extend to a larger distance for the decreasing density profile case A1 when compared to the increasing density profile case C1. However, if we consider the axial velocity component only, higher axial velocity continues to a larger distance for the increasing density profile case C1 than for the decreasing density profile case A1.
Both density profile cases A2 and C2 with $R_j = 4a$ where the CD kink is advected with the flow maintain more regular structure to larger distances than cases A1 and C1 with with $R_j = a/2$ where the CD kink is not advected with the flow.  

\subsection{Quantitative Kink and Flow Structure}

In order to reveal the helical structure more quantitatively, we show 2D slices made in the $x - z$ plane at $y=0$ in Figures 5 and 6.  The slices show the density, the azimuthal magnetic field component with azimuthal magnetic field magnitude contours, the azimuthal velocity component with azimuthal magnetic field magnitude contours, and the axial velocity component with axial magnetic field magnitude contours. 

Figure 5 shows the decreasing density cases A1 with $R_j = a/2$ and A2 with $R_j=4a$. For case A1 both density and axial velocity slices show that the jet flow is strongly distorted as a result of non-linear growth of the CD kink instability at $z < 5L = 20a = 40 R_j$. 
\begin{figure}[h!]
\epsscale{0.80}
\plotone{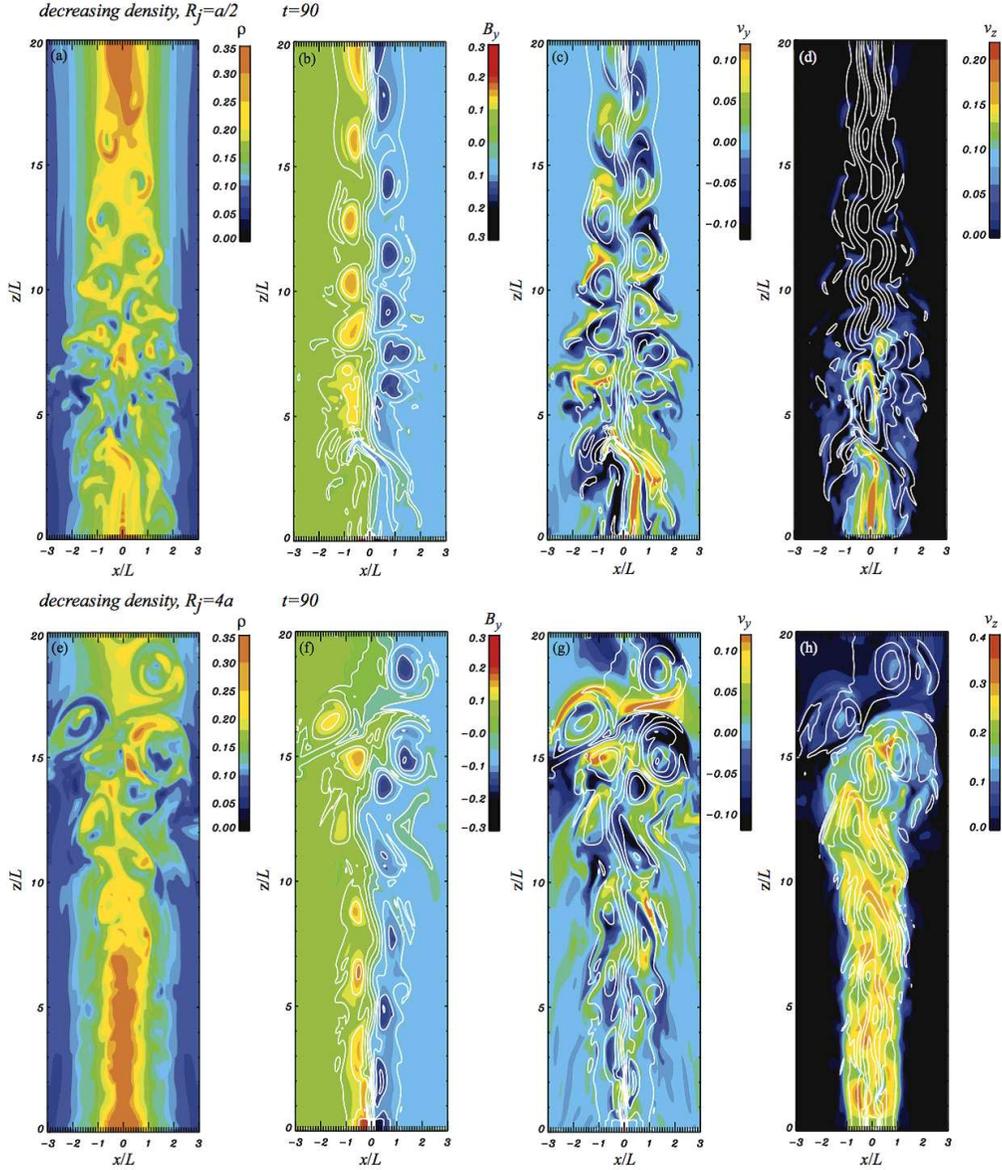}
\caption{2D slices at $t=90$ of  (a \& e) the density $\rho$, (b \& f) the azimuthal $B_y$ magnetic field component with $|B_y|$ magnitude contours, (c \& g) the azimuthal $v_y$ velocity component with $|B_y|$ magnitude contours, and (d \& h) the axial $v_z$ velocity component with axial $|B_z|$ magnetic field component magnitude contours. Slices are in the $x-z$ plane at $y=0$ for the decreasing density cases A1 with $R_j = a/2$ ({\it top panels}) and A2 $R_j = 4a$ ({\it bottom panels}). Note that the color scales for $v_z$ are different for case A1 and case A2. \label{decd_vyvz_roby_2D}} 
\end{figure}
Within the innermost region at $z < 2.5L = 10a = 20 R_j$ the azimuthal velocity slice shows rapid rotation of the flow that appears as $v_y \sim \pm 0.1c$ at $x/L \sim \pm 0.3$, and the axial velocity slice indicates acceleration to flow speeds $v_z > 0.2c$. Axial flow at speeds $v_z \sim 0.2c$ persists out to $z \lesssim 7.5L = 30a = 60R_j$. At larger distances from the inlet the azimuthal component of the magnetic field indicates temporal CD kink development with a wavelength of $\lambda_k^l \sim 2L$ that can also be seen in the $|B_z|$ contours. The initial axial flow appears primarily as azimuthal flow for $z > 7.5L$ that accompanies the helical magnetic field filament that winds around the jet plasma column at these larger distances. For example, the flow appears just below the $|B_y|$ maxima at $z/L \sim 11.5$, $z/L \sim 13$ and $z/L \sim 13$, as $v_y > 0$ at $z/L \sim 10.5$, $x/L \sim 0.7$ and $v_y < 0$ at $z/L \sim 12$, $x/L \sim -0.7$ and again as $v_y > 0$ at $z/L \sim 14.5$, $x/L \sim 0.7$,  respectively. 

For case A2 both density and axial velocity slices show that the jet maintains its initial collimation out to $z \sim 10L = 40a = 10 R_j$, and axial flow at speeds $v_z \gtrsim 0.2c$ persists out to $z \sim 15L = 60a = 15R_j$ albeit with considerable internal structure. The azimuthal velocity slice indicates helical flow inside $R_j$, but with the azimuthal component increasing along the jet to $|v_y| \gtrsim 0.1c$ at $z \sim 10L$. This spatial behavior of the azimuthal velocity is indicative of spatial growth of a propagating CD kink. 
The axial flow shows values up to $v_z \lesssim 0.35c$ even very close to the inlet. In general, the highest speeds are located at $|x/L| \sim L/2 = 2a = R_j/2$ inside the velocity shear surface.
For case A2 the azimuthal and axial magnetic field contours indicate CD kink development with a wavelength of $\lambda_k^l \sim 3L$.

Figure 6 shows the increasing density cases C1 with $R_j = a/2$ and C2 with $R_j=4a$. For case C1 both density and axial velocity slices indicate less helical distortion and slightly less twisted flow out to  $z \lesssim 7.5L = 30a = 60R_j$ than for the comparable decreasing density case A1. 
\begin{figure}[h!]
\epsscale{0.80}
\plotone{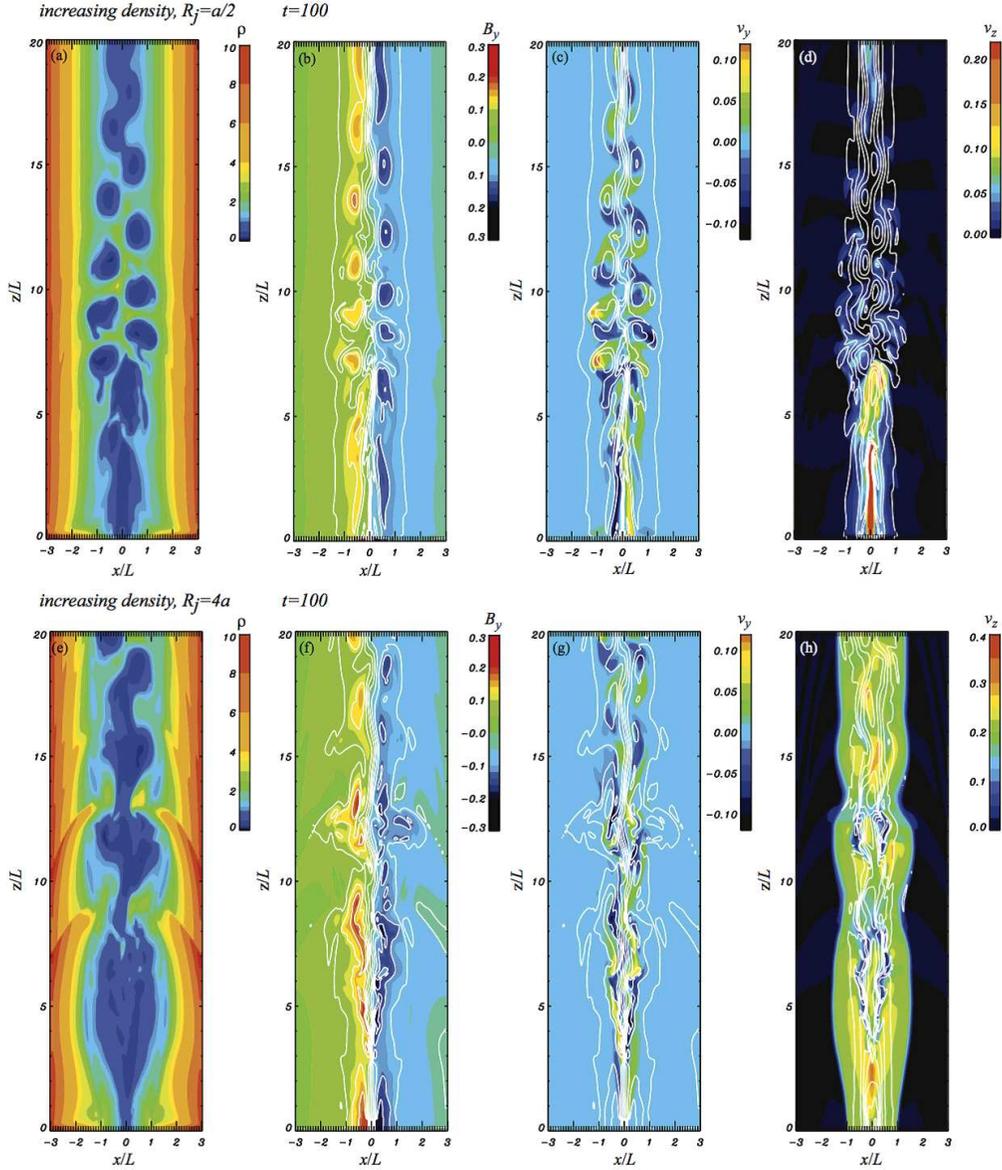}
\caption{2D slices at $t=90$ of  (a \& e) the density $\rho$, (b \& f) the azimuthal $B_y$ magnetic field component with $|B_y|$ magnitude contours, (c \& g) the azimuthal $v_y$ velocity component with $|B_y|$ magnitude contours, and (d \& h) the axial $v_z$ velocity component with axial $|B_z|$ magnetic field component magnitude contours. Slices are in the $x-z$ plane at $y=0$ for the decreasing density cases C1 with $R_j = a/2$ ({\it top panels}) and C2 $R_j = 4a$ ({\it bottom panels}). Note that the color scales for $v_z$ are different for case C1 and case C2. \label{incd_vyvz_roby_2D}} 
\end{figure}
Within the innermost region at $z < 4L = 16a = 32 R_j$ the azimuthal velocity slice shows rapid rotation of the flow also seen for case A1 that appears in the $v_y$ slice as $v_y \lesssim 0.1c$ at $x/L \sim 0.3$ and $v_y \gtrsim -0.1c$ at $x/L \sim -0.3$. 
The axial velocity slice indicates acceleration to flow speeds $v_z > 0.2c$ close to the inlet that was also seen for case A1. 
Axial flow at speeds $v_z \sim 0.2c$ persists out to  $z \sim 7.5L$ and appears less helically twisted than for the comparable decreasing density case A1. This behavior is confirmed by re-examination of the magnetic field line and velocity isosurface structure shown in Figure 4 panels (a) and (c) which also indicate a much more disordered in magnetic and flow field structure for the decreasing density case A1 at  $z < 7.5L$. On the other hand, at larger distances from the inlet the azimuthal and axial magnetic field contours indicate CD kink development with a wavelength of $\lambda_k^l \sim 2L$ that is the same as for case A1. Overall, case C1 maintains a more regular and less disordered structure on the computational grid than case A1.  This more regular structure is associated with reduced kink amplitudes in case C1 relative to case A1.

For case C2 both density and axial velocity slices indicate that the jet maintains its initial collimation across the computational grid, albeit with considerable internal and now some external structure. The azimuthal and axial velocity slices indicate twisted flow inside $R_j$. The CD kink appears to grow spatially with the
azimuthal velocity component increasing to $|v_y| \sim 0.1c$ at $z \sim 9L$, while the axial flow shows values up to $v_z \lesssim 0.35c$.  In general, the highest speeds are located at $|x| \sim L/2 = 2a = R_j/2$ inside the velocity shear surface.  Lack of spatial growth beyond $z \sim 9L$ out to about $z \sim 12L$ suggests CD kink growth saturation but firm confirmation requires both a longer simulation run time and a larger grid in the transverse direction to minimize transverse boundary effects.  For case C2 the azimuthal component of the magnetic field indicates CD kink development with a wavelength of $\lambda_k^l \sim 3L$ similar to  the wavelength found for case A2.  Overall, the flow and magnetic fields for case C2 are much more ordered than for case A2.  Additionally, the CD kink now propagates sufficiently faster than the external medium magnetosonic speed and drives shocks in the external medium.  The shocks indicated best by the density slice image are approximately axisymmetric. This axisymmetry is also indicated in the axial velocity slice and the 3D velocity isosurface in Figure 4d. We caution that this axisymmetry could be an artifact of the proximity of the transverse boundaries. We might expect a super-magnetosonic propagating helically twisted CD kink to produce a more or less continuous helically twisted shock along the forward edge of the helix. A much larger grid in the transverse direction is necessary to resolve this issue.

Additional information on the internal structure of the CD kink is provided by 1D cuts parallel to the jet axis in the $x-z$ plane at different radial, $x$, locations. In Figure 7, 1D cuts for case A1 with $R_j = a/2$ are made at $x = 0$, $a/2$, \& $2a$ and for case A2 with $R_j = 4a$ are made at $x = 0$, $2a$, \& $6a$. 
\begin{figure}[h!]
\epsscale{0.85}
\plotone{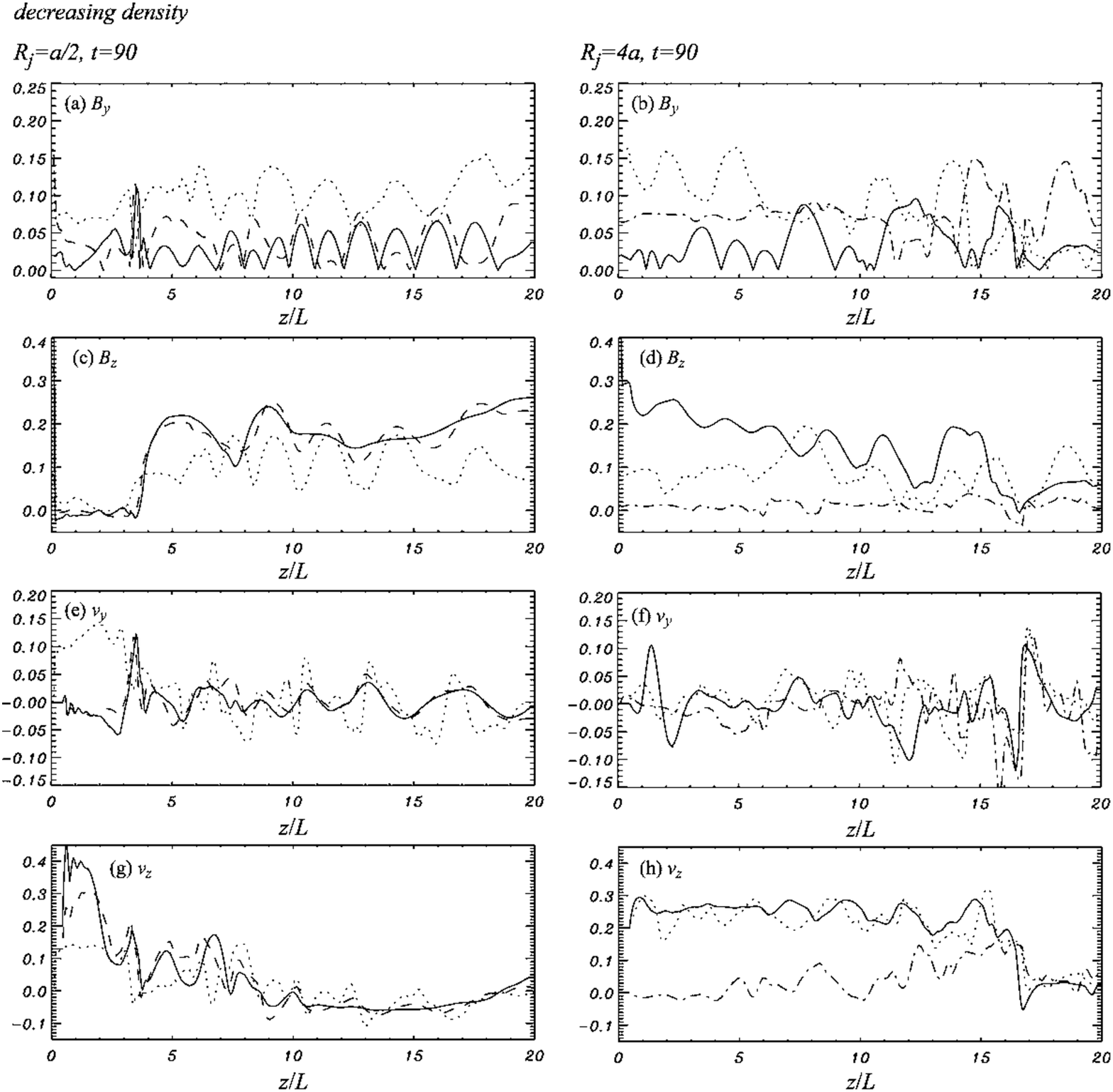}
\caption{1D cuts parallel to the jet axis of the azimuthal ($B_y$, $v_y$) and axial ($B_z$, $v_z$) components of the magnetic field and velocity located at $x=0$ (solid), $L/8=a/2$ (dashed), $L/2 =2a$ (dotted), and $3L/2=6a$ (dash-dotted) for the increasing density case A1 with $R_j=a/2$ ({\it left panels}) and A2 with $R_j=4a$ ({\it right panels}) at $t=90$. \label{dec_1D}}
\end{figure}
The 1D cuts reveal considerable azimuthal magnetic field structure associated with the CD kink in the azimuthal, $B_y$, component when $x \le 2a$ for both cases A1 and A2.
There is almost no significant magnetic field structure associated with the CD kink when $x > 2a$ for case A1 (not shown) and no significant structure associated with the CD kink when $x > R_j = 4a$ for case A2 (see the 1D cut at $x = 6a$). Both cases show similar oscillating structure associated with the CD kink along the 1D cut at $x = 2a$ albeit at the different wavelengths, (A1) $\lambda_k^l\sim 2L$ and (A2) $\lambda_k^l \sim 3L$ confirming the results seen in the 2D slices shown in Figure 5. Temporal development of the CD kink in both 2D slices and 1D cuts  show that the kink grows in place for case A1 and propagates with $v_k \sim v_z =  0.2c$ for case A2. Interestingly, both cases show CD kink structure at a shorter wavelength,  (A1)  $\lambda_k^s \sim 1.2L$ for all $z$ and (A2) $\lambda_k^s \sim 1.3L$ when $z < 7.5L$, along the axis, $x = 0$, that is not revealed in the 2D slices.  This internal shorter wavelength kink is confined to near the axis and exists only for $x \le a/2$. The spatial growth that occurs in case A2 obscures this shorter wavelength at larger distances from the inlet and leads to a much less regular kink structure in the 1D cuts at both wavelengths.  The longer wavelength CD kink can also be seen in the oscillations in the axial magnetic field component, $B_z$, for case A1 at $x = a/2~\&~ 2a$ and for case A2 at $x=2a$. Some indication of the azimuthal motions associated with the CD kink can be found in the $v_y$ velocity component along all 1D cuts in case A1 and in the 1D cut at $x=2a$ in case A2.  The axial, $v_z$, velocity component also shows some oscillating structure associated with the CD kink but is most useful in showing the distance, (A1) $z \sim 7L$ and (A2) $z \sim 16.5L$, reached by the jet flow.  This confirms our estimates based on the 3D velocity isosurfaces and the 2D velocity slices. 

In Figure 8, 1D cuts for case C1 with $R_j = a/2$ are made at $x = 0$, $a/2$, \& $2a$ and for case C2 with $R_j = 4a$ are made at $x = 0$, $2a$, \& $6a$.
\begin{figure}[h!]
\epsscale{0.85}
\plotone{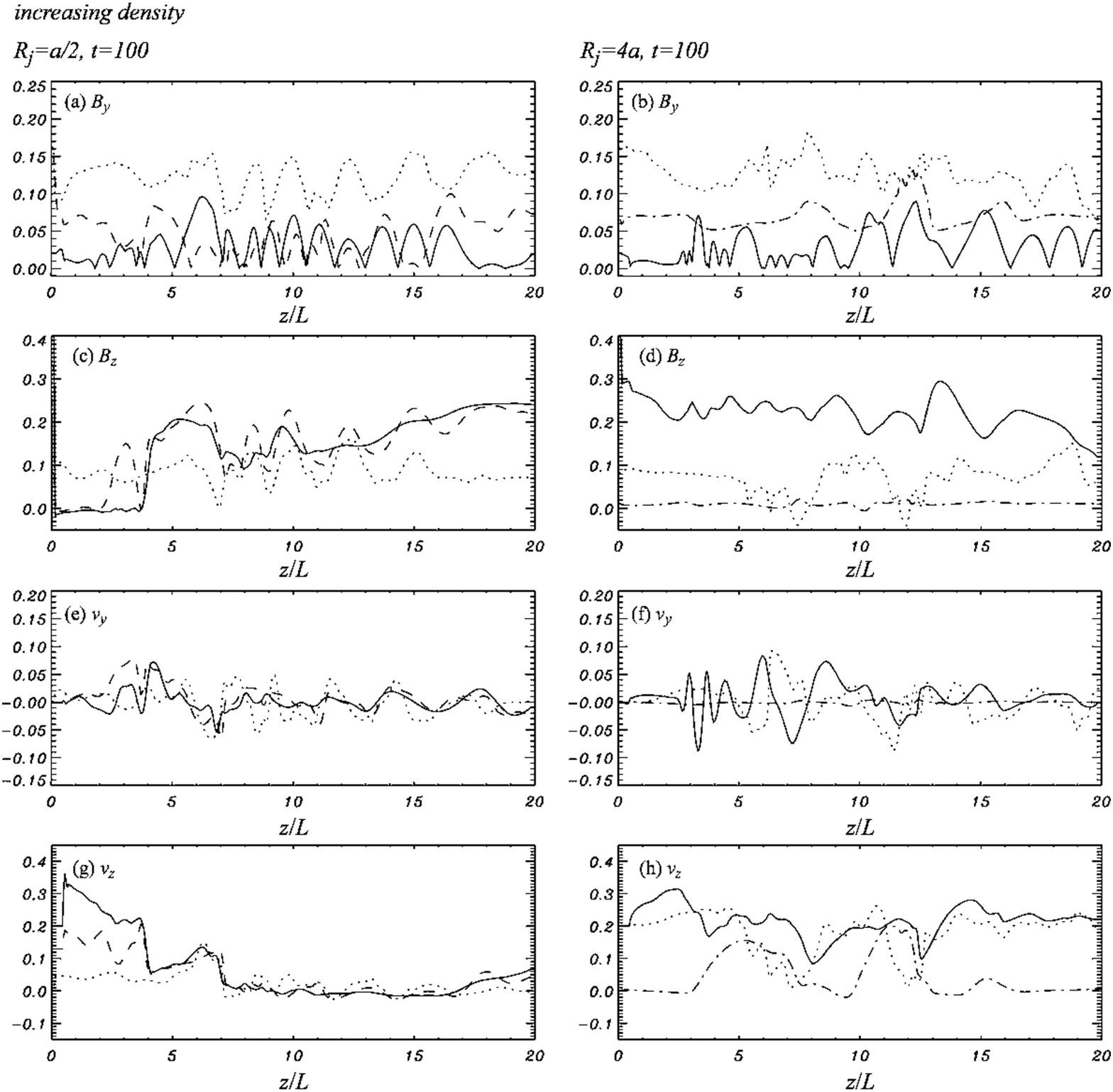}
\caption{1D cuts parallel to the jet axis of the azimuthal ($B_y$, $v_y$) and axial ($B_z$, $v_z$) components of the magnetic field and velocity located at $x=0$ (solid), $L/8=a/2$ (dashed), $L/2 =2a$ (dotted), and $3L/2=6a$ (dash-dotted) for the increasing density case C1 with $R_j=a/2$ ({\it left panels}) and C2 with $R_j=4a$ ({\it right panels}) at $t=100$. \label{inc_1D}}
\end{figure}
The 1D cuts reveal regular oscillating structure in the azimuthal, $B_y$, component associated with the CD kink when $x \le 2a$ for case C1 but much less  regular structure for case C2. There is almost no significant magnetic field structure associated with the CD kink when $x > 2a$ for case C1 (not shown) and no significant structure associated with the CD kink when $x > R_j = 4a$ for case C2 (see the 1D cut at $x = 6a$). The oscillating structure associated with the CD kink along the 1D cut at $x = 2a$ for case C1 has a wavelength $\lambda_k^l \gtrsim 2L$ similar to that found for case A1. While much less regular, the 1D cut at $x = 2a$ for case C2 shows oscillating structure consistent with the wavelength $\lambda_k^l  \sim 3L$ found for case A2 and confirms the results seen in the 2D slices shown in Figure 6. Temporal development of the CD kink in both 2D slices and 1D cuts  show that the kink grows in place for case C1 and propagates with $v_k \sim v_z =  0.2c$ for case C2. Similar to what was found for cases A1 and A2, cases C1 and C2 show CD kink structure at a shorter wavelength, (C1)  $\lambda_k^s \sim 1.1L$ for $7L < z < 16L$ and (C2) $\lambda_k^s \sim 2L$ for $z > 10L$, along the axis, $x = 0$, that is not revealed in the 2D slices.  This internal shorter wavelength kink is confined to near the axis and exists only for $x \le a/2$.   The longer wavelength CD kink can also be seen in the oscillations in the axial magnetic field component, $B_z$, for case C1 at $x = a/2$ \& $2a$ but not for case C2. Oscillations in $B_z$ for case C2 along the 1D cut at $x=0$ appear to correspond to the oscillations in $B_y$ along the 1D cut at $x=0$. Some indication of the azimuthal motions associated with the CD kink can be found in the $v_y$ velocity component along 1D cuts at $x=0$, $a/2$, \& $2a$ in case C1 and in the 1D cuts at $x=0$ \& $2a$ in case C2.  The axial, $v_z$, velocity component also shows some oscillating structure associated with the CD kink but is most useful in showing the distance, (C1) $z \sim 7L$ and (C2) $z > 20L$, reached by the jet flow.  This confirms our more qualitative estimates based on the 3D velocity isosurfaces and the 2D velocity slices. 

The 1D cuts show an axial acceleration to maximum speeds $v_z^{max} \sim 0.3 - 0.4c$ near to the inlet for all cases. For cases A1 and C1 with $R_j = a/2$, as indicated by the 2D slices, the acceleration is most evident along the axis. Azimuthal motion with $v_y \sim 0.1 -0.15c$ in the 1D cut at $x = 2a$ can clearly be seen for case A1. Note that this is outside the velocity shear radius. Azimuthal motion at a lower level appears in 1D cuts for case C1 but recall that values comparable to case A1 are indicated by the 2D slices. For cases A2 and C2 with $R_j = 4a$, as indicated by the 2D slices, axial acceleration occurs for $x \le 2a$ inside the velocity shear radius. Azimuthal motion associated with spatial growth of the CD kink, as indicated by the 2D slices, is seen in the 1D cuts at $x = 0~ \& ~2a$ for cases A2 and C2. 

We have verified that the observed axial acceleration and rotation near to the inlet are a result of the reconfiguration of the magnetic field induced by the precessional perturbation. We note that no significant acceleration or rotation of the initial steady non-rotating force-free configuration occurs at a specific location along the jet until after the precessional perturbation has passed that location, e.g., about half the simulation duration halfway down the jet. Temporal study at a fixed distance from the inlet shows that precession winds up the field amplifying the toroidal field component and reducing the axial component. $\mathbf{J \times B}$ forces associated with the resulting non-force free configuration appear in the axial and azimuthal directions. Some indication of this process near to the inlet can be seen in the 1D cuts shown in Figures 7 and 8.  The rapid decline in $B_y$ and $B_z$ at $z < 4a = L$ and subsequent slower decline in case A2 at $z < 3L$ occurs on the same length scale over which acceleration in $v_z$ occurs. Radial cuts (not shown) for cases A2 and C2 at $z = 12a = 3L$ reveal an axial velocity minimum at $x \lesssim 2a$, recall that $R_j = 4a$, and an associated reversal in azimuthal motion inside and outside this axial velocity minimum induced by $\mathbf{J \times B}$ forces. We note that the induced azimuthal motion inside and outside the axial minimum is in the opposite sense in cases A2 and C2. Lack of sufficient numerical resolution inside $R_j = a/2$ prevents detection of this type of behavior in cases A1 and C1.

RMHD simulations of jet production from an inner solid body and outer Keplarian rotating magnetosphere threaded by a poloidal magnetic field performed by Porth (2013) develop a rotating velocity profile that has some features similar to our results. In Porth (2013), the rotation is shown induced by the comoving reference frame Lorentz, $\mathbf{J \times B}$, and electric, $\rho_e \mathbf{E}$, forces (see Figure 10 in Porth 2013) where in general, the comoving pitch parameter, $P$, declines to some minimum outside the main jet and then increases. Faster rotation occurs outside the main jet where $P$ is smallest and the toroidal magnetic field is higher. Although in our simulations, we do not introduce jet rotation, the precessional perturbation winds up the initially force-free magnetic field whose observer reference frame $\mathbf{J \times B}$ forces induce rotation along with the observed axial acceleration.

\section{Kink Structure, Temporal and Spatial Growth}

We can compare the present simulations made on a non-periodic computational grid to previous temporal growth simulations made on a periodic computational grid.  In the previous periodic grid simulations it was found that a growing kink led to significant modification of the initial configuration after about ten e-folding times and the results, $\tau_{nl} \equiv 10 \tau_e$ are listed in Table 2 for declining density periodic grid simulations CP0 (static), CPsa/2 ($R_j = a/2$)  and CPs4a ($R_j = 4a$) from Mizuno et al.\ (2011) that are comparable to the present declining density simulations A1 and A2. Temporal growth of CPsa/2 on the periodic grid was similar to that found for a the static plasma column case CP0. Table 2 also contains an estimate of the time, $\tau_{nl}$, based on 3D and 2D images at which the present simulations show the first indication of non-linear behaviour. The time to non-linear behavior for comparable simulations, i.e., CPsa/2 and A1, and CPs4a and A2, are similar.  As the jet density declines relative to the external density, i.e., for cases B1 and B2 with constant density and for cases C1 and C2 with increasing density, $\tau_{nl}$ increases. The Lorentz factor of the flow is $\gamma_j \simeq$~1.02 and should have a negligible effect on the temporal and spatial growth rate of the CD kink instability which depends primarily on the Alfv\'{e}n velocity (e.g., Appl et al. 2000) but also depends on relativistic time dilation.  

The location of the  velocity shear radius has similar consequences for kink propagation, $v_k^{obs}$, in previous temporal growth simulations using a periodic computational grid and the present simulations using a non-periodic computational grid.  For a velocity shear surface well inside the characteristic radius, i.e. $R_j = a/2$, the kink does not propagate in the previous temporal growth simulations or in the present simulations. For the velocity shear surface well outside the characteristic radius, i.e., $R_j = 4a$, the kink propagates at about the flow speed in both the previous and present simulations. Note that in case CPs4a the range in kink propagation speeds indicated in Table 2 reflects a slow down in the observed kink propagation speed as the kink amplitude increased. This effect now shows up in A2 as the flow and kink propagation slow beyond $z \sim 12L$ (see panel h in Figure 7).

The speed, $v^{obs}_{pr}$, at which the initial perturbation travels across the computational grid in the present simulations is also given in Table 2. The slowest observed speed, $v^{obs}_{pr} = 0.22$c, associated with case C1 requires a time $t/t_c \sim 90$ for the precessional perturbation to cross the computational grid.  This is about twice the time, $\tau_{nl}/t_c \sim 45$, required for non-linear behavior to develop in case C1. We find that the observed speed at which the initial perturbation travels across the grid can approximately be given by an average magnetosonic speed projected along the $z$-axis combined with the jet speed at the characteristic magnetic radius $a$ where
$$
v^{cal}_{pr} \equiv {{\langle v_{ms} \rangle_z + v_j(a)}\over{1 + \langle v_{ms} \rangle_z v_j(a)/c^2}}
$$
and $\langle v_{ms} \rangle_z \equiv \left[v_{ms}(a)\sin \theta_a + v_{ms}(R_j)\sin \theta_{R_j}\right] /2$ where for constant pitch $\theta_{R_j} = \theta_a \equiv \tan^{-1} |B_{\phi}/B_z|_a = 45\arcdeg$ is evaluated at the characteristic magnetic radius $a$.  Values for $v_{ms}$ at $R_j =$ a/2, a, \& 4a can be found in Table 1. Thus, the initial perturbation propagates as a magnetosonic wave in cases A1, B1 and C1 where $R_j = a/2$ and $v_j(a) = 0$, and as a magnetosonic wave advected with the flow in cases A2, B2 and C2 where $R_j = 4a$ and $v_j(a) = 0.2$c.

For the previous periodic computation grid simulations the wavelength of the kink, $\lambda^l_k = 12a$, was enforced by the length of the computational grid. Note that according to the Kruskal-Shafranov criterion, the instability develops at $\lambda_k > 2\pi a$ and a shorter kink wavelength $\lambda^s_k = 6a$ at half the length of the computational grid was not seen on the periodic grid as it was likely stable. For the static case of constant pitch and uniform density Appl et al. (2000) found a fastest growing wavelength $\lambda^*_k \simeq 8.4a$ and a corresponding growth time of $\tau^*_e \simeq 7.5 a/ v_{A0}$.  For a moving kink we might expect this temporal growth rate in the lab frame to be reduced with $\tau_e \propto \gamma_k$ where $\gamma_k$ is the moving kink Lorentz factor, e.g., Narayan et al. (2009).
 
In simulations A1, B1 and C1 we find a temporally growing kink wavelength of $\lambda^l_k \sim 2L = 8a$ that is on the order of the fastest growing wavelength $\lambda^* \sim 1.33 \lambda_{min} \sim 8.4 a$, predicted by a linear stability analysis for constant magnetic pitch where $\lambda_{min} = 2\pi P_c = 2\pi a$ is the shortest unstable wavelength (Bodo et al. 2013 and see Figure 1 in Bodo et al. 2013). In our simulations constant pitch $P_c \equiv RB_z/B_{\phi} = a$ as $B_z = B_{\phi}$ at $R = a$ (see Figure 1b). A shorter amplitude saturated kink wavelength, $\lambda^s_k \simeq$ (A1) $4.8a$, (B1) $5.0a$ and (C1) $4.4a$, is found to be operating in the inner part of the jet where $R \le a/2$. The longer wavelength kink amplitude grows most rapidly near to the inlet and the flow and magnetic structure become disordered relatively close to the inlet as a result of non-linear growth of the longer kink wavelength. 
 
In simulations A2, B2 and C2 we find a spatially growing kink wavelength of $\lambda^l_k \sim 3L = 12a = \lambda_j = 2\pi v_j/\omega_j$ that is at about the wavelength given by wave propagation at the jet speed associated with the perturbation frequency. This wavelength is about 40\% longer than the predicted fastest growing wavelength.  Again a shorter amplitude saturated kink wavelength, $\lambda^s_k \simeq$ (A2) $5.2a$, (B2) $5.8a$ and (C2) $8.0a$, is found to be operating in the inner part of the jet where $R \le a/2$. Distortion of the initial configuration is associated with non-linear growth of the longer wavelength. 

In all cases a shorter kink wavelength develops in the inner part of the jet. With the exception of case C2, the shorter kink wavelength is less than the minimum predicted unstable wavelength associated with constant magnetic pitch $P_c = a$.  1D radial cuts show that reconfiguration of the magnetic and flow field induced by the precessional perturbation reduces the magnetic pitch in the inner part of the jet from $P_c =a$ to $P \lesssim a$ at $R \sim a/2$ to $P \sim 0.4 a$ on the jet axis. This decrease in the pitch when $R < a/2$ allows shorter wavelengths to become unstable in the inner part of the jet. We note that in case C2 the pitch declined minimally in the inner part of the jet and this explains why the shorter kink wavelength in this case was at about the fastest growing wavelength associated with $P \sim P_c = a$.  

The development of an inner shorter wavelength is similar to what was found in rapidly rotating velocity shear periodic grid simulations (Mizuno et al.\, 2012). In those simulations an inner shorter unstable wavelength was associated with a reduced pitch parameter, $P$, accompanying faster jet rotation cases, i.e., see Figure 7 and Figure 8 for cases alp1om4 with $\Omega_0 = 4$ and alp1om6 with $\Omega_0 = 6$ in Mizuno et al. (2012). These cases showed an inner shorter unstable wavelength where the pitch parameter $P$ was smaller and an outer longer unstable wavelength where the pitch parameter $P$ was larger.  

Quantitative comparison between our results and instability predictions is difficult because no general spatial stability analysis has been performed for magnetically dominated jets.  
We found previously that for constant pitch cases the characteristic time for non-linear behavior was roughly $\tau_{nl} \sim 10 \tau_e$, with values for $\tau_e$ being dependent on the structure of the undisturbed state. In the present jet context the growing kink remains static or propagates with the flow depending on the location of the velocity shear surface. Comparison between the time required to achieve non-linear behavior  in the present simulations (determined qualitatively by examination of temporal development of 1D cuts and 2D slices) and the characteristic time to non-linear behavior in previous simulations (see Table 2) is similar and roughly in agreement with temporal stability analysis. 

In spatial growth cases an approximate distance to non-linear behavior at time $\tau_{nl}$ is (A2) $\sim 6L = 24a$, (B2) $\sim 7.5L = 30a$ and (C2) $\sim 9L = 36a$ and is consistent with $\ell_{nl} \sim \tau_{nl} v_k$ with $v_k \sim v_j$. This distance to non-linear behavior is maintained for $t > \tau_{nl}$. We can make some comparison between CPs4a (Mizuno et al. 2011) and A2. In CPs4a the initial axisymmetric structure was strongly distorted by the kink instability even though not disrupted and the predicted distance to non-linear behavior for CPs4a of $\ell_{nl} = \tau_{nl} v_k \sim 30a$ and the observed non-linear distance for A2 of $\ell_{nl} \sim 24a$ is in rough agreement. In order to check whether the instability disrupts jet flow, one has to compare $\ell_{nl}$ with a distance $\ell_{fl}$ to which high speed axial flow at $\simeq v_j$ continues at the end of the simulations.  The distance $\ell_{fl}$ at which $v_z < 0.2 c$ as determined from 2D slices at $t/t_c = 90 - 100$ is given in Table 2. This distance to flow disruption is more than twice the non-linear spatial growth distance estimate of  $\ell_{nl} \sim \gamma^{\alpha}_k v_k \tau_{nl}$ with $\gamma_k \sim 1.02$ and $\alpha = 3$ suggested by the temporal growth and kink propagation found in our periodic grid simulations (Mizuno et al.\ 2011). Here $\alpha=1$ corresponds to time dilation only and $\alpha = 3$ to time dilation plus inertial effects from a relativistically moving fluid.

\begin{deluxetable}{cccccccccc}
\tablecolumns{10}
\tablewidth{0pc}
\tablecaption{e-foldings, speeds, lengths \tablenotemark{\,*}}
\label{table2}
\tablehead{
\colhead{Case} & \colhead{$v_{j}/c$} & \colhead{$R_{j}/a$} & \colhead{$\tau_{nl}/t_{c}$\tablenotemark{\,a}} & \colhead{$v^{obs}_{pr}/c$} & \colhead{$v^{cal}_{pr}/c$} & \colhead{$v^{obs}_k/c$} & \colhead{$\lambda_k^l/a$} & \colhead{$\lambda_k^s/a$} & \colhead{$\ell/a$\tablenotemark{\,b}}
} 
\startdata
CP0  & 0.0 & 0.0 & 37.5 & --- & --- & 0.0 & 12 & --- & 0 \\
CPsa/2 & 0.2 & 0.5 & 37.5 & --- & --- & $\sim 0$ & 12 & --- & 0 \\
CPs4a & 0.2 & 4.0 & 40.5   & --- & --- & 0.20 - 0.16 & 12 & --- & 32 - 26 \\
\tableline
A1  & 0.2 & 0.5 & $\sim 35$ & 0.29 $\pm$ 0.01 & 0.28 & $\sim 0$ & $\sim 8$ & 4.8 & $\sim12$ \\
A2  & 0.2 & 4.0 & $\sim 35$ & 0.39 $\pm$ 0.01 & 0.42 & $\sim 0.2$ & $\sim12$ & 5.2 & $\sim 60$ \\
B1  & 0.2 & 0.5 & $\lesssim 40$ & 0.26 $\pm$ 0.01 & 0.24 & $\sim 0$ & $\sim 8$ & 5.0 & $\sim 16$ \\
B2  & 0.2 & 4.0 & $\lesssim 40$ & 0.37 $\pm$ 0.01 & 0.37 & $\sim 0.2$ & $\sim12$ & 5.8 & $ > 75$ \\
C1  & 0.2 & 0.5 & $\sim 45$ & 0.22 $\pm$ 0.01 & 0.23 & $\sim 0$ & $\sim 8$ & 4.4 & $\sim 25$ \\
C2  & 0.2 & 4.0 & $\sim 45$ & 0.33 $\pm$ 0.01 & 0.31 & $\sim 0.2$ & $\sim 12$ & 8.0 & $> 80$ \\
\enddata
\tablenotetext{*}{~A, B \& C cases observed values $\pm 10\%$ unless otherwise noted.}
\tablenotetext{a}{~CP cases: $\tau_{nl} \equiv 10 \tau_e$ using $\tau_e$ from Mizuno et al.\ (2011) ; $t_c \equiv 4a/c$}
\tablenotetext{b}{~CPs4a case: $\ell = \ell_{nl}  \equiv \tau_{nl} v_k$ ; A, B, \& C cases: $\ell = \ell_{fl} \equiv \ell(v_z < 0.2c)$ @ $t/t_c = 90 - 100$. }
\end{deluxetable}

\clearpage

\section{Summary and Discussion}

We have investigated the influence of velocity shear location and radial density profile on the spatial development of the CD kink instability using a non-periodic computational box. In these spatial growth simulations and the previous temporal growth simulations (Mizuno et al.\ 2011) the mildly relativistic jet is established across the computational domain with a helical force-free magnetic field. Our previous study of temporal growth of the CD kink instability used a periodic computational box and uniform spatial perturbation of the jet across the box that only allowed for temporal development of wavelengths that fit within the box. The kink propagated or was static depending on the velocity shear radius relative to the characteristic radius of the magnetic field. For the case of constant helicity and constant magnetic pitch that we have considered here, the characteristic radius corresponds to the radius at which the toroidal magnetic field is a maximum.  For temporal growth cases studied previously using a periodic computational box (Mizuno et al. 2011) with increasing or decreasing helicity, i.e., for decreasing or increasing magnetic pitch, the characteristic radius is outside or inside the radius at which the toroidal magnetic field is a maximum, respectively, but not by a large amount. 

Our temporal studies (Mizuno et al.\ 2011) showed that results are strongly affected by the choice of velocity shear radius with respect to the characteristic radius of the helical magnetic field. Simulations found a temporally growing static kink for a velocity shear radius of $R_j = a/2$ that is the same as the velocity shear radius of our present cases (A1, B1, C1). Kink propagation was found to increase as the velocity shear radius increased and found a temporally growing kink propagating at about the jet speed for a velocity shear radius of $R_j = 4a$ that is the same as the velocity shear radius of our present cases (A2, B2, C2). Other temporal studies investigated the influence of jet rotation and differential motion on the CD kink instability (Mizuno et al.\ 2012). The most unstable of these temporal simulations developed both longer and shorter kink wavelengths at larger and smaller jet radii, respectively, and this combination led to disruption of the initial cylindrical configuration.  The shorter wavelength inner kink developed when the wavelength at one-half the periodic computational box length became unstable as a result of increasing the magnetic helicity (decreasing the magnetic pitch) in the inner portion of the jet. While this decrease in the magnetic pitch was associated with jet rotation, in general the temporal growth properties were not much different than for the comparable non-rotating static cases studied in Mizuno et al. (2009).

In the present simulations, we find that the CD kink instability is partially stabilized by a radially increasing density structure. Current observations and GRMHD simulation of jet formation suggest that an inner high speed jet is surrounded by a denser and slower outer flow (e.g., McKinney 2006). Therefore, we expect that in the GRMHD simulation by McKinney \& Blandford (2009), a CD kink develops in the jet but saturates in part as a result of radially increasing density structure. This stabilizing effect helps us to understand the existence of non-destructive kink structure in observed relativistic jets. The kink (wiggle) structure is developed by the CD instability near to the jet origin but is sufficiently stabilized at larger distances by the radially increasing density structure and other effects such as jet expansion, axial velocity and rotation profiles, etc. and we will investigate these effects in future work.

Our simulations indicate that CD kink wavelength(s) and propagation dependent on the location of velocity shear, magnetic pitch profile and characteristic magnetic radius. A characteristic magnetic radius inside the velocity shear combined with a less dense jet inside a denser confining medium is stabilizing. In order to make any comparison with observed helical structures we need to perform radiation transfer calculations based on simulation results making image and polarization maps to directly compare with observations. Ultimately the combination of simulation generated images and correspondingly detailed observation images might allow the determination of AGN jet radial velocity and magnetic field profiles. Existing VLBI observations indicate moving helical jet structures suggesting an effective velocity shear radius on the order of or greater than the characteristic radius of the magnetic field. To further resolve this issue would require VLBI resolution sufficient to make a comparison between polarization observations giving radial magnetic structure and proper motion observations giving radial velocity structure. Such futuristic observations could be used to pin down where the velocity is most strongly sheared compared to the characteristic radius of the jet's magnetic field.

In this study, we have used an initially constant axial speed with a non-rotating but CD kink unstable setup as a first step in our investigation of spatial growth of the CD kink instability. The precessional perturbation that we impose at the inlet winds up the initially non-rotating force-free magnetic field configuration leading to jet acceleration and rotation. Interestingly, we find a reversal in the sense of rotation across an axial velocity minimum that is well inside the velocity shear radius. Such a reversal in rotation is not seen in the propagating jet simulations performed by Porth (2013) although his simulations have shown faster rotation accompanying the location of stronger toroidal magnetic fields wound up by the inlet rotation profile. These profile differences show, as found by O'Neill et al.\ (2012), that structural details are significantly dependent on the initial configuration. A more realistic setup for relativistic jets would involve more physically relevant radial toroidal and poloidal velocity and magnetic profiles along with jet acceleration and expansion. Jet acceleration and expansion change the behavior of the instabilities seen on jets modeled as cylinders of constant radius. Jet expansion has a spatially stabilizing effect as spatial growth scales with the characteristic magnetic and jet radius. Additionally, magnetic instabilities become ineffective when the Alfv\'{e}n speed drops below the lateral expansion speed of the jet. Currently only a few MHD simulations have explored the effect of jet expansion on the development of the CD kink instability (Moll et al. 2008; Moll 2009; Porth 2013). In future work, we plan to investigate the effect of jet acceleration and expansion on the development of CD kink instability in the relativistic MHD regime along with more realistic radial profiles.   

We have used a mildly relativistic jet speed, $v_j=0.2c$, in the simulations. This has the advantage of allowing  comparison to non-relativistic 3D MHD simulations and will provide a baseline for comparison with relativistic jet speed simulations. In general, our present results are similar to those obtained from non-relativistic 3D MHD simulations of jet formation and propagation (e.g., Nakamura \& Meier 2004; Moll et al. 2008; Moll 2009). For example, exponential growth times in our simulations and previous non-relativistic MHD simulations are generally on the order of Alfv\'{e}n crossing times. In non-relativistic MHD simulations of jet propagation, the helically twisted structures developed by CD kink instability move with the jet propagation speed (e.g., Nakamura \& Meier 2004; Moll et al. 2008; Moll 2009), and we find similar results provided the jet's velocity shear radius is outside the characteristic radius of the helical magnetic field. The advantage to our setup, unlike jet generation and propagation based simulations, is control of radial and axial profiles including acceleration and expansion. This will allow a systematic determination of the most important stabilizing factors.

In our current and past simulation work, we assume ideal MHD with no resistivity other than induced by the numerical scheme. Thus, relativistic magnetic reconnection (e.g., Blackman \& Field 1994; Lyubarsky 2005; Watanabe \& Yokoyama 2006; Zenitani et al. 2010b; Takahashi et al. 2011; Zanotti \& Dumbser 2011; Baty et al. 2013; Takamoto 2013) associated with non-linear kink development is not treated. Development of resistive relativistic MHD (RRMHD) codes are now one of the frontiers of relativistic MHD (e.g., Komissarov 2007; Palenzuela et al. 2009; Dumbser \& Zanotti 2009; Zenitani et al. 2010b; Takahashi et al. 2011; Takamoto \& Inoue 2011). We have developed an RRMHD code (Mizuno 2013). The present and ongoing work using ideal MHD will be used to establish a baseline for comparison with CD kink instability work allowing for relativistic magnetic reconnection.

\acknowledgments
This work has been supported by NSF awards AST-0908010 and AST-0908040 to UA and UAH,  and NASA awards NNX08AG83G and NNX12AH06G to UAH. Y.M. acknowledges support from Taiwan National Science Council award NSC 100-2112-M-007-022-MY3. The simulations were performed on the Pleiades Supercomputer at the NAS Division of the NASA Ames Research Center, the SR16000 at YITP in Kyoto University, and the Kraken at the National Institute for Computational Sciences in the XSEDE project supported by National Science Foundation.


\begin{thebibliography}{}

\bibitem[Aloy \& Rezzolla(2006)]{Alo06} Aloy, M. A., \& Rezzolla, L. 2006, \apjl, 640, L115

\bibitem[Appl et al.(2000)]{App00} Appl, S., Lery, T., \& Baty, H. 2000, \aap, 355, 818

\bibitem[Balbus \& Hawley(1998)]{Bal98} Balbus, S. A., \&  Hawley, J. F. 1998, Rev. Mod. Phys., 70, 1

\bibitem[Bateman(1978)]{Bat78} Bateman, G. 1978, MHD instabilites (Cambridge, Mass., MIT Press, 1978, p.270)

\bibitem[Baty et al.(2013)]{Bat13} Baty, H., Petri, J., \& Zenitani, S. 2013, \mnras, 436, L20

\bibitem[Beckwith \& Stone(2011)]{Bec11} Beckwith, K., \& Stone, J. M. 2011, \apjs, 193, 6

\bibitem[Beckwith et al.(2008)]{Bec08} Beckwith, K., Hawley, J. F., \& Krolik, J. H. 2008, \apj, 678, 1180

\bibitem[Begelman(1998)]{Beg98} Begelman, M. C. 1998, \apj, 493, 291

\bibitem[Begelman et al.(1980)]{Beg80} Begelman, M. C., Blandford, R.D., \& Rees, M.J. 1980, \nat, 287, 307

\bibitem[Begelman et al.(2008)]{Beg08} Begelman, M. C., Fabian, A. C., \& Rees, M. J. 2008, \mnras, 384, L19

\bibitem[Beskin \& Nokhrina(2006)]{Bes06} Beskin, V. S. \& Nokhrina, E. E. 2006, \mnras,  367, 375

\bibitem[Blackman \& Field(1994)]{Bla94} Blackman, E. G., \& Field, G. B. 1994, \prl, 72, 494

\bibitem[Blandford(1976)]{Bla76} Blandford, R. D. 1976, \mnras, 176, 465

\bibitem[Blandford(2000)]{Bla00} Blandford, R. D. 2000, Phil. Trans. Roy. Soc. A, 358,

\bibitem[Blandford \& Znajek(1977)]{Bla77} Blandford, R. D. \& Znajek, R. L. 1977, \mnras, 179, 433

\bibitem[Bodo et al.(2013)]{Bod13} Bodo, G., Mamatsashvili, G., Rossi, P., \& Mignone, A. 2013, \mnras, 434, 3030

\bibitem[Carey \& Sovinec(2009)]{Car09} Carey, C.S. \& Sovinec, C.R. 2009, \apj, 699, 362

\bibitem[De Villiers et al.(2003)]{DeV03} De Villiers, J.-P., Hawley, J. F.,
\& Krolik, J. H. 2003, \apj, 599, 1238

\bibitem[De Villiers et al.(2005)]{DeV05} De Villiers, J.-P., Hawley, J. F.,
Krolik, J. H., \& Hirose, S. 2005, \apj, 620, 878

\bibitem[Del Zanna(2007)]{Del07} Del Zanna, L., Zanotti, O., Bucciantini, N., \& Londrillo, P. 2007, \aap, 473, 11

\bibitem[Drenkhahn(2002)]{Dre02a} Drenkhahn, G. 2002, \aap, 387, 714

\bibitem[Drenkhahn \& Spruit(2002)]{Dre02b} Drenkhahn, G. \& Spruit, H. C. 2002, \aap, 391, 1141

\bibitem[Dumbser \& Zanotti(2009)]{Dum09} Dumbser, M., \& Zanotti, O. 2009, J. Comput. Phys., 228, 6991

\bibitem[Eichler(1993)]{Eic93} Eichler, D. 1993, \apj, 419, 111

\bibitem[Ghisellini \& Tavecchio(2008)]{Ghi08} Ghisellini, G., \& Tavecchio, F. 2008, \mnras, 386, L28

\bibitem[Giannios \& Spruit(2006)]{Gia06} Giannios, D., \& Spruit, H. C. 2006, \aap, 450, 887

\bibitem[Giannios et al.(2009)]{Gia09} Giannios, D., Uzdensky, D. A., \& Begelman, M. C. 2009, \mnras, 395, L29

\bibitem[G\'{o}mez et al.(2001)]{Gom01} G\'{o}mez, J. L., Guirado, J. C., Agudo, I., Marscher, A. P., Alberdi, A., Marcaide, J. M., \& Gabuzda, D. C. 2001, \mnras, 328, 873

\bibitem[Granot(2012a)]{Gra12a} Granot, J. 2012a, \mnras, 421, 2442

\bibitem[Granot(2012b)]{Gra12b} Granot, J. 2012b, \mnras, 421, 2467

\bibitem[Granot et al.(2011)]{Gra11} Granot, J., Komissarov, S. S., \& Spitkovsky, A. 2011, \mnras, 411, 1323

\bibitem[Guan et al.(2013)]{Gua13} Guan, X., Li, H., \& Li, S. 2013, \apj, in press (arXiv:1312.0661)

\bibitem[Hardee(2004)]{Har04} Hardee, P.E. 2004, \apss, 293, 117

\bibitem[Hardee(2007)]{Har07} Hardee, P.E. 2007, \apj, 664, 26

\bibitem[Hardee et al.(2007)]{Har07a} Hardee, P., Mizuno, Y., \& Nishikawa, K.-I. 2007, \apss, 311, 283

\bibitem[Hawley \& Krolik(2006)]{HK06} Hawley, J. F. \& Krolik, J. H. 2006, \apj, 641, 103

\bibitem[Istomin \& Pariev(1994)]{IP94} Istomin, Y. N. \& Pariev, V. I. 1994, \mnras, 267, 629

\bibitem[Istomin \& Pariev(1996)]{IP96} Istomin, Y. N. \& Pariev, V. I. 1996, \mnras, 281, 1

\bibitem[Komissarov(1997)]{Kom97} Komissarov, S. S. 1997, Phys. Lett. A, 232, 435

\bibitem[Komissarov(2007)]{Kom07} Komissarov, S. S. 2007, \mnras, 382, 995

\bibitem[Komissarov \& Barkov(2009)]{Kom09} Komissarov, S. S., \& Barkov, M. V. 2009, \mnras, 397, 1153

\bibitem[Komissarov et al.(2010)]{Kom10} Komissarov, S. S., Vlahakis, N., \& K\"{o}nigl, A. 2010, \mnras, 407, 17

\bibitem[Komissarov et al.(2007)]{Kom07} Komissarov, S. S., Barkov, M. V., Vlahakis, N., \& K\"{o}nigl, A. 2007, \mnras, 380, 51

\bibitem[Komissarov et al.(2009)]{Kom09} Komissarov, S. S., Vlahakis, N., K\"{o}nigl, A., \& Barkov, M. V. 2009, \mnras, 394, 1182

\bibitem[Lery et al.(2000)]{LB00} Lery, T., Baty, H., \& Appl, S. 2000, \aap, 355, 1201

\bibitem[Levinson(2007)]{Lev07} Levinson, A. 2007, \apjl, 671, L29

\bibitem[Lobanov \& Zensus(2001)]{Lob01} Lobanov, A. P. \& Zensus, J. A. 2001, Science, 294, 128

\bibitem[Lovelace(1976)]{Lov76} Lovelace, R. V. E. 1976, \nat, 262, 649

\bibitem[Lyubarskii(1992)]{Lyu92} Lyubarskii, Y. E. 1992, Sov. Astr. Lett., 18, 356

\bibitem[Lyubarskii(1999)]{Lyu99} Lyubarskii, Y. E. 1999, \mnras, 308, 1006

\bibitem[Lyubarsky(2005)]{Lyu05} Lyubarsky, Y. E. 2005, \mnras, 358, 113

\bibitem[Lyubarsky(2009)]{Lyu09} Lyubarsky, Y. 2009, \apj, 698, 1570

\bibitem[Lyubarsky(2010a)]{Lyu10a} Lyubarsky, Y. E. 2010a, \mnras, 402, 353

\bibitem[Lyubarsky(2010b)]{Lyu10b} Lyubarsky, Y. 2010b, \apj, 725, L234

\bibitem[Lyubarsky(2011)]{Lyu11} Lyubarsky, Y. 2011, \pre, 83, 016302

\bibitem[Lyutikov(2011)]{Lyut11} Lyutikov, M. 2011, \mnras, 411, 422

\bibitem[McKinney(2006)]{McK06} McKinney, J. C. 2006, \mnras, 368, 1561

\bibitem[McKinney \& Blandford(2009)]{McK09} McKinney, J. C., \& Blandford, R. D., 2009, \mnras, 394, L126

\bibitem[McKinney \& Gammie(2004)]{McK04} McKinney, J. C., \& Gammie, C. F. 2004, \apj, 611, 977

\bibitem[McKinney \& Uzdensky(2012)]{McK12} McKinney, J. C., \& Uzdensky, D. A. 2012, \mnras, 419, 573

\bibitem[Mignone et al.(2010)]{Mig10} Mignone, A., Tzeferacos, P., \&  Bodo, G. 2010, J. Comput. Phys., 229, 5879

\bibitem[Mizuno(2013)]{Miz13} Mizuno, Y. 2013, \apjs, 205, 7

\bibitem[Mizuno et al.(2007)]{Miz07} Mizuno, Y., Hardee, P., \& Nishikawa, K.-I. 2007, \apj, 662, 835

\bibitem[Mizuno et al.(2011)]{Miz11} Mizuno,Y., Hardee, P. E., \& Nishikawa, K.-I. 2011, \apj, 734, 19

\bibitem[Mizuno et al.(2009)]{Miz09} Mizuno, Y., Lyubarsky, Y., Nishikawa, K.-I., \& Hardee, P. E. 2009, \apj, 700, 684

\bibitem[Mizuno et al.(2012)]{Miz12} Mizuno, Y., Lyubarsky, Y., Nishikawa, K.-I., \& Hardee, P. E. 2012, \apj, 757, 16

\bibitem[Mizuno et al.(2008)]{Miz08} Mizuno, Y., Hardee, P., Hartmann, D. H., Nishikawa, K.-I., \& Zhang, B. 2008, \apj, 672, 72

\bibitem[Mizuno et al.(2006)]{Miz06} Mizuno, Y., Nishikawa, K.-I., Koide, S., Hardee, P., \& Fishman, G. J. 2006, preprint, (arXiv:astro-ph/0609004)

\bibitem[Moll(2009)]{Mol09} Moll, R. 2009, \aap, 507, 1203

\bibitem[Moll et al.(2008)]{Mol08} Moll, R., Spruit, H. C., \& Obergaulinger, M. 2008, \aap, 492, 621

\bibitem[Nakamura \& Meier(2004)]{Nak04}  Nakamura, M., \& Meier, D. L. 2004, \apj, 617, 123

\bibitem[Nakamura et al.(2007)]{Nak07} Nakamura, M., Li, H., \& Li, S. 2007, \apj, 656, 721

\bibitem[Narayan et al.(2009)]{Nar09} Narayan, R., Li, J., \& Tchekhovskoy, A. 2009, \apj, 697, 1681

\bibitem[O'Neill et al.(2012)]{ONe12} O'Neill, S. M., Beckwith, K., \& Begelman, M. C. 2012, \mnras, 422, 1436

\bibitem[Ouyed et al.(2003)]{Ouy03} Ouyed, R., Clarke, D.A., \& Pudritz, R.E. 2003, \apj, 582, 292

\bibitem[Palenzuela et al.(2009)]{Pal09} Palenzuela, C., Lehner, L., Reula, O., \& Rezzolla, L. 2009, \mnras, 394, 1727

\bibitem[Penna et al.(2010)]{Pen10} Penna. R. F., McKinney, J. C., Narayan, R., Tchekovskoy, A., Shafee, R., \& McClintock, J. E. 2010, \mnras, 408, 752

\bibitem[Porth(2013)]{Por13} Porth, O. 2013, \mnras, 429, 2428

\bibitem[Sikora et al.(2005)]{Sik05} Sikora, M., Begelman, M. C., Madejski, G. M., \& Lasota, J.-P. 2005, \apj, 625, 72

\bibitem[Spruit et al.(2001)]{Spr01} 	Spruit, H. C., Daigne, F., \& Drenkhahn, G. 2001, \aap, 369, 694

\bibitem[Spruit et al.(1997)]{Spr97} Spruit, H. C., Foglizzo, T., \& Stehle, R. 1997, \mnras, 288, 333

\bibitem[Takahashi et al.(2011)]{Tak11} Takahashi, H. R., Kudoh, T., Masada, Y., \& Matsumoto, J. 2011, \apjl, 739, L53

\bibitem[Takamoto(2013)]{Taka13} Takamoto, M. 2013, \apj, 775, 50

\bibitem[Takamoto \& Inoue(2011)]{Taka11} Takamoto, M., \& Inoue, T. 2011, \apj, 735, 113

\bibitem[Tchekhovskoy et al.(2009)]{Tch09} Tchekhovskoy, A., McKinney, J. C., \& Narayan, R. 2009, \apj, 699, 1789

\bibitem[Tchekhovskoy et al.(2010)]{Tch10} Tchekhovskoy, A., Narayan, R., \& McKinney, J. C. 2010, New Astronomy, 15, 749

\bibitem[Tomimatsu et al.(2001)]{Tom01} Tomimatsu, A., Matsuoka, T., \& Takahashi, M. 2001, \prd, 64, 123003

\bibitem[Vlahakis \& K\"{o}nigl(2003)]{Vla03} Vlahakis, N., \& K\"{o}nigl, A. 2003, \apj, 596, 1080

\bibitem[Watanabe \& Yokoyama(2006)]{Wat06} Watanabe, N., \& Yokoyama, T. 2006, \apjl, 647, L123

\bibitem[Zanotti \& Dumbser(2011)]{Zan11} Zanotti, O., \& Dumbser, M. 2011, \mnras, 418, 1004

\bibitem[Zenitani et al.(2010a)]{Zen10a} Zenitani, S., Hesse, M., \& Klimas, A. 2010a, \apj, 712, 951

\bibitem[Zenitani et a.(2010b)]{Zen10b} Zenitani, S., Hesse, M., \& Klimas, A. 2010b,  \apjl, 716, L214

\end{thebibliography}
\end{document}